\documentclass[english,english, aps, preprint, superscriptaddress]{revtex4-1}
\usepackage[T1]{fontenc}
\usepackage[latin9]{inputenc}
\usepackage{textcomp}
\usepackage{graphicx}
\usepackage{esint}

\makeatletter

\providecommand{\tabularnewline}{\\}

\@ifundefined{textcolor}{}
{%
 \definecolor{BLACK}{gray}{0}
 \definecolor{WHITE}{gray}{1}
 \definecolor{RED}{rgb}{1,0,0}
 \definecolor{GREEN}{rgb}{0,1,0}
 \definecolor{BLUE}{rgb}{0,0,1}
 \definecolor{CYAN}{cmyk}{1,0,0,0}
 \definecolor{MAGENTA}{cmyk}{0,1,0,0}
 \definecolor{YELLOW}{cmyk}{0,0,1,0}
 }

\makeatother

\usepackage{babel}
\begin{document}

\preprint{preprint}

\title{Transport properties of RTILs from classical molecular dynamics}

\author{Oliviero Andreussi}

\email{oliviero.andreussi@epfl.ch}

\affiliation{Massachusetts Institute of Technology, Department of Material Science
and Engineering, 77 Massachusetts Ave. Cambridge 02139 MA, USA}

\affiliation{Theory and Simulations of Materials, École Polytechnique Fédérale
de Lausanne, Station 12, 1015 Lausanne, Switzerland}

\author{Nicola Marzari}

\email{nicola.marzari@epfl.ch}

\affiliation{Theory and Simulations of Materials, École Polytechnique Fédérale
de Lausanne, Station 12, 1015 Lausanne, Switzerland}
\begin{abstract}
Room Temperature Ionic Liquids (RTILs) have attracted much attention
in the scientific community in the past decade due the their novel
and highly customizable properties. Nonetheless their high viscosities
pose serious limitations to the use of RTILs in practical applications.
To elucidate some of the physical aspects behind transport properties
of RTILs, extensive classical molecular dynamics (MD) calculations
are reported. Bulk viscosities and ionic conductivities of butyl-methyl-imidazole
based RTILs are presented over a wide range of temperatures. The dependence
of the properties of the liquids on simulation parameters, e.g. system
size effects and choice of the interaction potential, is analyzed.
\end{abstract}
\maketitle

\section{introduction}

In recent years, room temperature ionic liquids (RTILs) have see much
interest due to their promising properties in {}``gree chemistry''
applications \cite{plechkova_chemsocrev_2008}. Similarly to the well
known high temperature molten salts, RTILs are liquids composed solely
by ions. As opposite to molten salts, the presence of large asymmetric
organic cations inhibits crystallization and allows these salts to
be liquid at temperature as low as 100\textdegree{}C. Despite the
fact that the first report of a RTIL dates back to the beginning of
the last century \cite{walden_bais_1914}, the first generation of
RTILs, based on chloroaluminate(III) systems, was theoretically predicted
\cite{wilkes_fsrl_1982} and experimentally realized \cite{wilkes_inorgchem_1982,hussey_pac_1988}
just in the 1980's. Nonetheless, practical applications of chloroaluminate
ionic liquids were strongly limited by their high moisture sensitivity
\cite{abdulsada_chemcomm_1987,abdulsada_dalton_1993}. It is just
with the advent of water and air stable RTILs \cite{wilkes_chemcomm_1992}
that the full potential of these new compounds became apparent. Several
reviews on this topic appeared in the last years \cite{weingartner_acie_2008,plechkova_chemsocrev_2008,binnemans_chemrev_2007,dupont_chemrev_2002,greaves_chemrev_2008,hapiot_chemrev_2008,vanrantwijk_chemrev_2007},
clearly highlighting the interest in this field .

Properties of RTILs are as different from standard molecular solvents
as ionic crystals differ from molecular ones. Due to the ionic nature
of their constituents, RTILs generally show negligible vapor pressure
and high thermal stability, and they tend to be very good solvent
for most organic and inorganic species. Moreover, they typically present
very high electrochemical stability and are intrinsically able to
conduct electrical currents. Despite all these remarkable properties,
none of the existing RTILs possess all of them at the same time. Nonetheless,
since RTILs are based on organic molecules, they can be easily modified
by standard chemical reactions. Indeed, both the cation and the anion
can be individually modified in order to tune the physico-chemical
properties of the resulting RTIL. This allows a great degree of flexibility
in the design of the most suitable compound, given that the number
of different possible combinations scales as the product of the number
of oppositely charged ions. When considering the number of binary
and ternary mixtures of all available cations and anions, the range
of possibilities quickly diverges. In this perspective, the idea of
task-specific RTILs has emerged and applications of RTILs in the most
diverse fields have been proposed, and patented. Organic synthesis
and catalysis, extraction, and treatment of rare-earths elements are
some of the most studied fields of application of RTILs. 

Particular attention has been devoted to the possible applications
of RTILs in electrochemistry \cite{hapiot_chemrev_2008}. Specifically,
RTILs as electrolytes in dye-sensitized solar cells \cite{bonhote_inorgchem_1996,ito_naturephotonics_2008,kuang_acie_2008},
fuel cells\cite{fernicola_cpc_2007}, and lithium batteries \cite{fernicola_jpowersource_2007}have
been extensively studied both for safety and efficiency. Indeed, the
low volatility of these liquids insures their safety against combustion
and explosion. On the other hand, some of the existing RTILs present
very high electrochemical stability, thus allowing the use of higher
voltages in batteries and improving the overall efficiency of the
devices. A major limitation in actual applications of RTILs as electrolytes
is represented by the high viscosities of all RTILs known to date.
Even though the conductivity of some of the available RTILs is already
sufficiently high for battery applications \cite{fernicola_jpowersource_2007},
design rules to improve ionic conductivity are currently the subject
of intense experimental and theoretical study. In this context, much
of the scientific effort is directed toward understanding of the molecular
mechanisms that influence the transport properties of RTILs. 

Despite the huge increase in the literature on RTILs, experimental
results still suffer from some limitations, and in particular can
be affected by the presence of water and impurities in the systems
studied \cite{seddon_pac_2000}. Indeed, water was shown to strongly
affect transport properties of RTILs, with a decrease of diffusion
coefficients by orders of magnitude in water-RTILs mixtures being
reported \cite{rollet_jpcb_2007}. A full quantitative characterization
of the amount and kind of impurities is often lacking and results
can be difficult to compare. 

In order to provide a detailed picture of the molecular mechanisms
behind the properties of RTILs, several theoretical works have appeared
in the literature. The first semi empirical and ab-initio calculation
on RTILs were performed at the end of the 1980s \cite{wilkes_inorgchem_1982,wilkes_fsrl_1982},
with the main purpose of investigating the stability and the structure
of isolated ions; theoretical literature on RTILs started increasing
after 2001, when some of the first parametrization of classical interaction
potentials (force fields) were reported \cite{hanke_molphys_2001,morrow_jpcb_2002,shah_greenchem_2002,shah_fluidphaseeq_2004,margulis_jpcb_2002,deandrade_jpcb_2002,deandrade_jpcb_2002b}.
The first calculations typically also reported results on single ions
and ion pairs as obtained from density-functional theory (DFT). These
results were used to fit classical empirical potentials that were
exploited to compute some of the fundamental properties of the liquids.
In order to validate the force fields against available experimental
results, radial and angular distribution functions, densities and
diffusion coefficients were the first quantities investigated. The
agreement between simulation results and experimental data were not
always satisfactory, with some approximation in the functional form
of the interaction potential being too crude \cite{hunt_molsim_2006}.
In particular, it was shown that using an All Atom picture instead
of a Unified Atoms Method was more effective in reproducing experimental
densities \cite{shah_fluidphaseeq_2004}. Similarly, including polarization
effects was shown to significantly improve the agreement with experimental
results, in particular for transport properties \cite{yan_jcpb_2004}.
Despite some of the limitations of the available force fields, MD
simulations were effective in predicting and validating some of the
new features characteristic of the RTILs' structure, such as the presence
of heterogeneities and holes in the liquids \cite{hu_pnas_2007}.
Nonetheless, some properties still lack an accurate analysis and characterization.
This is particularly true for transport properties, such as viscosity
and conductivity, that, due to the slow glassy-like dynamics of most
RTILs, require long simulation times and large system sizes. Indeed,
while parametrized force fields are generally checked against structural
properties, few groups have reported careful analysis of viscosities
and conductivities \cite{habasaki_jcp_2008,qiao_jpcb_2008,cadena_jpcb_2006b,reycastro_jpcb_2006a,reycastro_jpcb_2006b}.
For these reasons, in the present work an extensive set of classical
MD simulations on prototypical RTILs are reported. After a short methodological
section (II), our results on the effect of temperature and simulation
parameters on the transport properties are presented (IIIA). Eventually,
results on ionic liquids based on different anions are compared in
Section IIIB.

\section{Simulation details}

Classical molecular dynamics (MD), as implemented in the DL\_POLY
program, was used throughout \cite{smith_jmolgraph_1996,smith_molsim_2002,smith_molsim_2006}.
Three different 3-butyl-1-methylimidazolium (BMIM) based RTILs were
examined, namely the salts obtained with the PF$_{6}^{-}$, BF$_{4}^{-}$,
and bis(trifluorometylsulfonyl)-imide (Tf$_{2}$N$^{-}$) anions.
As for the interaction potential, three different force field were
considered for the case of BMIM-PF$_{6}$ \cite{morrow_jpcb_2002,liu_jpcb_2004,lopes_jpcb_2004a,lopes_jpcb_2004c},
while simulations on BMIM-BF$_{4}$ and BMIM-Tf$_{2}$N were performed
using the force fields developed by Canongia-Lopes et al. \cite{lopes_jpcb_2004a,lopes_jpcb_2004b,lopes_jpcb_2004c}.
An interaction cutoff radius of 15 Å was used throughout. Initial
configurations containing 128, 432 and 1024 ion pairs were generated
starting from a fcc cubic lattice with the ions occupying random lattice
sites. Equilibration in the NPT ensemble was enforced using Berendsen's
thermostat and barostat \cite{berendsen_jcp_1984}, with equilibration
runs performed until convergence of the statistical average of the
density was achieved. Following equilibration, up to four different
production runs for each system were performed in the NVT ensemble,
using the Nose-Hoover thermostat \cite{nose_molphys_1984,hoover_pra_1985}.
A simulation timestep of 2 fs was used for temperatures lower or equal
to 500 K, while 1 fs was chosen for simulations at higher temperatures. 

Diffusion coefficients of the ions were computed in terms of mean
square displacements (MSD) 
\[
\left\langle r_{i}^{2}\left(t\right)\right\rangle =\left\langle \left|\vec{r}_{i}\left(t\right)-\vec{r}_{i}\left(0\right)\right|^{2}\right\rangle ,
\]
using the Einstein relation \cite{frenkel_book_2001,allen_book_1989}
\begin{equation}
D_{i}=\lim_{t\rightarrow\infty}\frac{1}{6}\frac{\partial\left\langle r_{i}^{2}\left(t\right)\right\rangle }{\partial t}.\label{eq: diffusion coefficient}
\end{equation}
Following standard methodology (e.g. as reported in \cite{cadena_jpcb_2006b}),
to assess the true diffusive behavior of the ions, MSDs will be displayed
in log-log plots, together with their slopes $\beta\left(t\right)$
as a function of time
\begin{equation}
\beta(t)=\frac{d\log MSD(t)}{d\log t}.\label{eq: beta}
\end{equation}
At very short times, a value of $\beta$ equal to two is expected,
corresponding to a free, ideally ballistic, motion of the ions. On
the other extreme, at very long times $\beta$ should reach a value
equal to one, corresponding to real diffusive regime, in which the
mean square displacement of the ion grows linearly with time. At intermediate
times a sub-linear behavior is expected, characterized by a logarithmic
slope $\beta(t)$ lower than one. We would like to stress here that
this analysis is crucial in determining the effective achievement
of diffusive behavior and to determine the portion of the MSD plot
that should be used to fit diffusion coefficients. This is particularly
important in the case of slow, viscous liquids such as RTILs. 

The Einstein formalism was also used in the calculation of ionic conductivities
\cite{frenkel_book_2001,allen_book_1989} 
\begin{equation}
\lambda=\lim_{t\rightarrow\infty}\frac{e^{2}}{6Vk_{B}T}\frac{\partial\left\langle \sum_{i,j=1}^{N}z_{i}z_{j}\vec{r}_{i}\left(t\right)\vec{r}_{j}\left(t\right)\right\rangle }{\partial t},\label{eq: conductivity}
\end{equation}
while viscosities ($\eta$) were computed from the auto correlation
of the off diagonal elements of the stress tensor $\sigma^{xy}$,
exploiting the Green-Kubo relation \cite{frenkel_book_2001,allen_book_1989}
\begin{equation}
\eta=\frac{V}{k_{B}T}\intop_{0}^{\infty}\left\langle \sigma^{xy}\left(0\right)\sigma^{xy}\left(t\right)\right\rangle \mathrm{d}t.\label{eq: viscosity}
\end{equation}

\section{Results}

\subsection{BMIM-PF6}

A system composed of 128 BMIM-PF6 ion pairs, described with the interaction
potential parametrized by Liu et al. (ref. \cite{liu_jpcb_2004},
abbreviated in the following as LHW2004), was equilibrated at several
temperatures, ranging from 300 K to 1000 K. Densities, reported in
Table \ref{tab: pf6_temperature}, are in agreement with what reported
in the literature and show a good match with the experimental results
\cite{tokuda_jpcb_2004,jin_jpcb_2008} available at the lowest temperatures.
\begin{table}
\begin{tabular}{|c|c|c|c|c|c|}
\hline 
Systems & $\rho$$\left(\mathrm{g}/\mathrm{c}\mathrm{m}^{3}\right)$ & $D^{+}\left(\mathrm{m}^{2}/\mathrm{s}\right)$ & $D^{-}\left(\mathrm{m}^{2}/\mathrm{s}\right)$ & $\eta\left(\mathrm{mPa\, s}\right)$ & $\lambda\left(\mathrm{S/cm}\right)$\tabularnewline
\hline 
\hline 
298.15 K (exp \cite{jin_jpcb_2008}) & $1.37$ & $1.6\pm0.2\cdot10^{-11}$ & - & $196\pm20$ & $4.04\pm0.4\cdot10^{-3}$\tabularnewline
\hline 
300 K (exp \cite{tokuda_jpcb_2004}) & $1.37$ & $8.0\pm3.0\cdot10^{-12}$ & $5.9\pm2.4\cdot10^{-12}$ & $230\pm65$ & $1.66\pm0.3\cdot10^{-3}$\tabularnewline
\hline 
300 K & $1.36$ & $4.8\pm1.0\cdot10^{-13}$ & $2.6\pm0.5\cdot10^{-13}$ & $2400\pm2000$ & $1.5\pm0.7\cdot10^{-4}$\tabularnewline
\hline 
350 K & $1.33$ & $4.0\pm1.0\cdot10^{-12}$ & $2.0\pm0.8\cdot10^{-12}$ & $940\pm400$ & $1.9\pm0.8\cdot10^{-3}$\tabularnewline
\hline 
375 K & $1.32$ & $1.0\pm0.1\cdot10^{-11}$ & $6.2\pm1.0\cdot10^{-12}$ & $421\pm200$ & $1.4\pm0.5\cdot10^{-3}$\tabularnewline
\hline 
400 K & $1.29$ & $3.1\pm0.1\cdot10^{-11}$ & $2.0\pm0.3\cdot10^{-11}$ & $111\pm4$ & $8.7\pm0.6\cdot10^{-3}$\tabularnewline
\hline 
450 K & $1.26$ & $1.14\pm0.06\cdot10^{-10}$ & $7.9\pm0.8\cdot10^{-11}$ & $31\pm1$ & $1.5\pm0.3\cdot10^{-2}$\tabularnewline
\hline 
500 K & $1.22$ & $2.5\pm0.2\cdot10^{-10}$ & $1.9\pm0.1\cdot10^{-10}$ & $12\pm1$ & $4.8\pm1.0\cdot10^{-2}$\tabularnewline
\hline 
600 K & $1.16$ & $8.5\pm0.1\cdot10^{-10}$ & $7.5\pm0.04\cdot10^{-10}$ & $7\pm1$ & $9.7\pm1.0\cdot10^{-2}$\tabularnewline
\hline 
1000 K & $0.87$ & $7.6\cdot10^{-9}$ & $7.3\cdot10^{-9}$ & $5$ & $2.25\cdot10^{-1}$\tabularnewline
\hline 
\end{tabular}\caption{\label{tab: pf6_temperature}Densities $\rho$, cation diffusion coefficients
$D^{+}$, anion diffusion coefficients $D^{-}$, viscosities $\eta$,
and conductivities $\lambda$ of the BMIM-PF6 RTIL as a function of
temperature, as computed by classical NVT molecular dynamics simulations
on a periodic cell of 128 ion pairs using the LHW2004 force field.}

\end{table}
\begin{figure}
\includegraphics[width=0.5\textwidth]{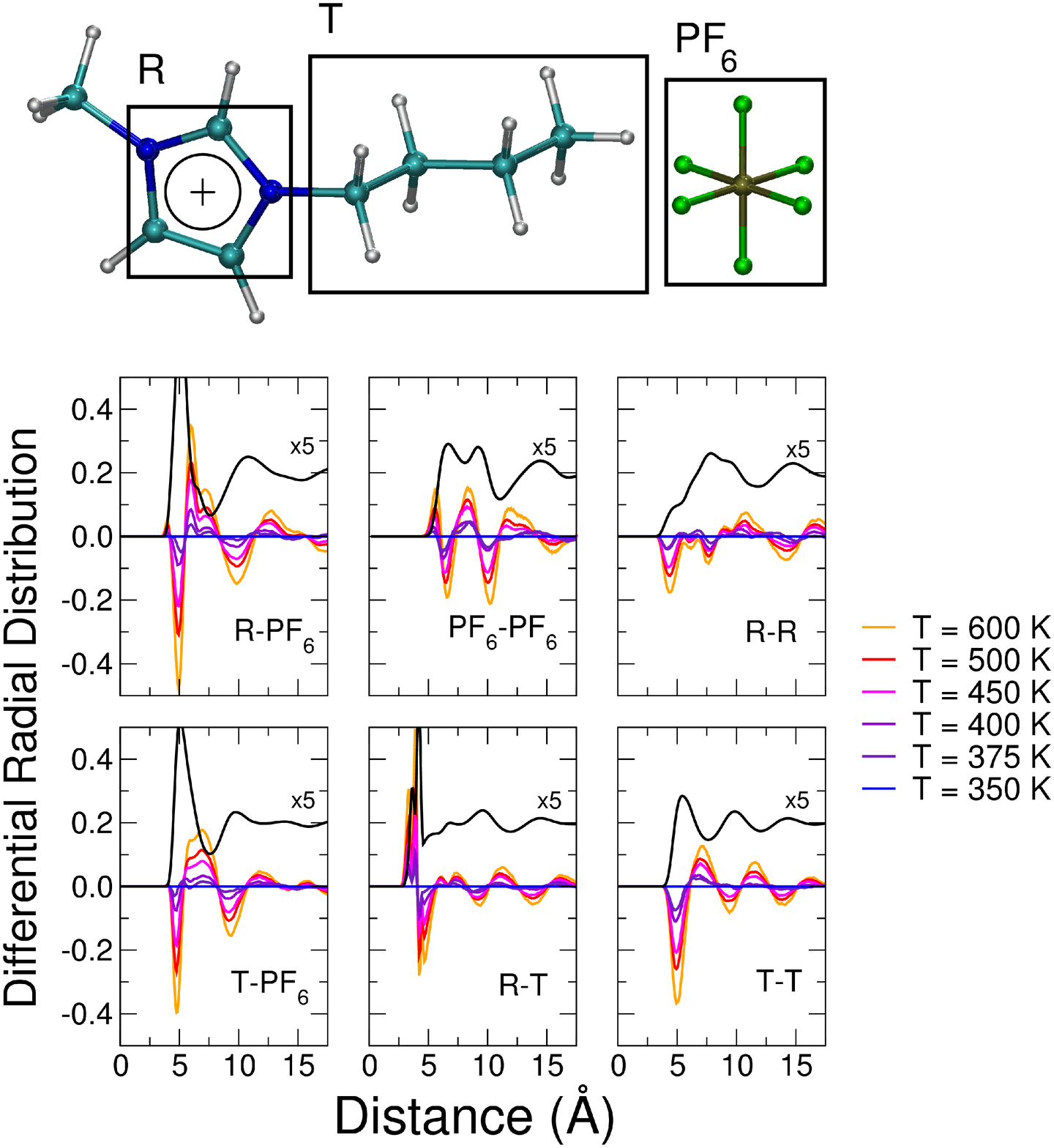}\caption{\label{fig: rdf_detailed_128ionpairs}Radial distribution functions
of BMIM-PF$_{6}$ RTIL, as computed by classical NVT molecular dynamics
simulations on a periodic cell of 128 ion pairs described by the LHW2004
force field. The BMIM cation (top panel) is partitioned in two distinct
subregions, corresponding to the charged aromatic ring (R) and the
long alkylic tail (T). The radial distribution functions at 300 K
are reported in black, while for higher temperatures the differentials
radial distributions with respect to 300 K have been reported (colored
lines). Strong charge induced correlations are evident from the leftmost
series of panels. A reasonably strong correlation is also present
between the alkyl tails of the cation, as shown in the T-T plot (rightmost
series, central panel). All the distribution functions show non-negligible
correlations for distances as long as half the cell size, even at
the highest temperatures.}
\end{figure}
The radial distribution functions of the system (reported in Figure
\ref{fig: rdf_detailed_128ionpairs}) are also in very good agreement
with previous results \cite{liu_jpcb_2004}. In order to provide a
more detailed description of the structure of the liquid, when analyzing
radial distributions the cation has been subdivided in two distinct
regions, corresponding to the charged aromatic ring (R) and the long
alkylic tail (T). Charge-induced correlations in the liquid can be
clearly evinced from the reported plots and persist, with some broadening,
even at the highest temperatures considered. Van der Waals interactions,
instead, are responsible for the high degree of correlation between
the alkyl chains of the cations. Due to the local, short-range nature
of the Van der Waals interaction, the first peak in the tail-tail
plot is one of the most sensible to temperature, shifting towards
larger distances as the simulation temperature increases.

\subsubsection{Transport Properties}

\begin{figure}
\includegraphics[width=0.5\textwidth]{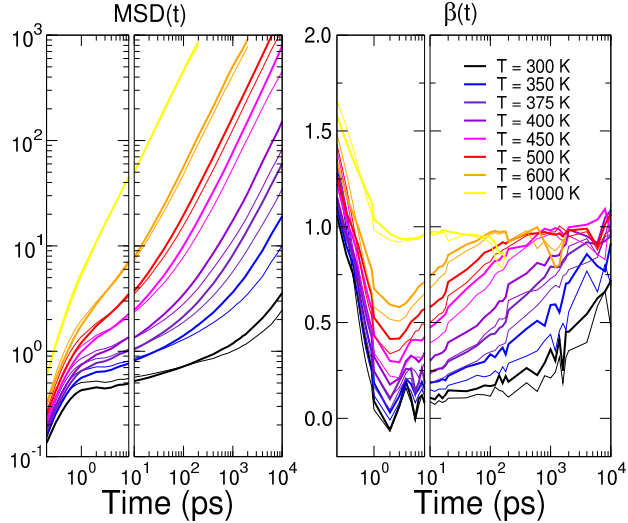}\caption{\label{fig: msd_detailed}Mean square displacements (left panel) and
corresponding values of $\beta$ (Eq. \ref{eq: beta}) for the BMIM-PF$_{6}$
RTIL, as computed by classical NVT molecular dynamics simulations
on a periodic cell of 128 ion pairs described by the LHW2004 force
field. Cations are represented by thick lines, while thin lines are
used for the anions. In order to characterize MSDs at short times,
a separate set of simulations was used, characterized by a very short
(0.5ps) time between sampling different configurations.}
\end{figure}

Mean square displacements of the different ions are reported in Figure
\ref{fig: msd_detailed} as a function of time, for the range of temperatures
studied.

\begin{figure}
\includegraphics[width=0.5\textwidth]{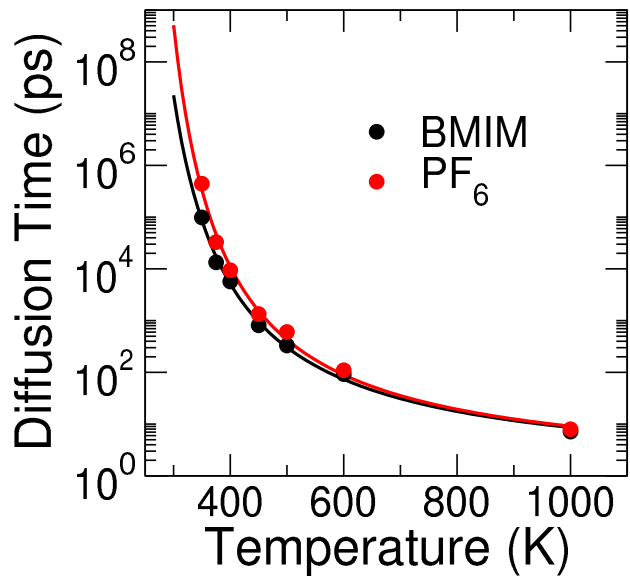}\caption{\label{fig: msd_times}Diffusion times, i.e. times necessary to the
ions to reach a diffusive regime with $\beta\simeq1$, as extracted
from classical NVT molecular dynamics simulations on a periodic cell
of 128 BMIM-PF$_{6}$ ion pairs, as described by the LHW2004 force
field (filled circles). Non-linear curve fits of the form $f=A\exp(B/(T-T_{0}))$
are added to the computed results and represented with solid lines.}
\end{figure}

 Despite the remaining noise in the logarithmic derivatives at long
times, it is seen from these plots that a diffusive behavior is reached
within the simulation times at each temperature but the lowest one.
Diffusion times, i.e. the times necessary to the ions to reach a diffusive
regime with $\beta\simeq1$, are summarized in Figure \ref{fig: msd_times}.
The curves show a divergent behavior at low temperature and can be
fitted with good accuracy by a Vogel-Fulcher-Tammann relation $f=A\exp(B/(T-T_{0}))$.
A value of $B\simeq1720/1730$ and a divergence temperature $T_{0}\simeq200/215\:\mathrm{K}$
can be extracted for the diffusion times of the cation and the anion
respectively. By extrapolating these behavior to the lowest temperature
considered ($300\:\mathrm{K}$), a diffusion time of the order of
$10^{8}$ ps can be estimated. Thus, a straightforward determination
of the diffusion coefficients of the ions at this temperature appears
beyond the limits of standard computational resources. For this reason,
the different transport properties of the system at $300\:\mathrm{K}$
are only estimated by averaging the results over several independent
calculations. In order to improve the statistics in the results and
get an estimate of the errors, the same approach was used for simulations
below $600\:\mathrm{K}$, that show diffusion times of the order of
few nanoseconds.

\begin{figure}
\includegraphics[width=0.5\textwidth]{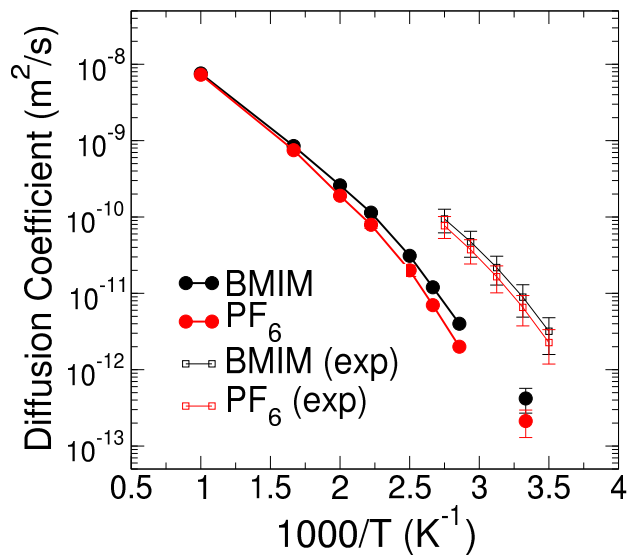}\caption{\label{fig: diffusion}Diffusion coefficients of cations (black) and
anions (red) as extracted from the MSDs computed on a periodic cell
of 128 BMIM-PF$_{6}$ ion pairs, as described by the LHW2004 force
field (filled circles). Experimental results from Ref. \cite{tokuda_jpcb_2004}
are also reported for comparison (empty squares). }
\end{figure}

Diffusion coefficients of both the cations and the anions, reported
in Figure \ref{fig: diffusion} and in Table \ref{tab: pf6_temperature},
show Arrhenius behavior at high temperatures, while a deviation from
linearity is evident at the lowest temperatures. As expected, diffusion
coefficients at $300\:\mathrm{K}$ appear overestimated, due to the
lack of a true diffusive regime in the simulations at this temperature.
 Activation energies of $E=31.4/28.4\cdot10^{-3}\mathrm{kJ}/\mathrm{mol}$
for the diffusion of the cation/anion can be inferred from the high
temperature slopes of the curves reported in Figure \ref{fig: diffusion}.
The change in slope at low temperature is an indication of the glassy
behavior of the liquid, consistent with the very slow non-ergodic
dynamics of the systems in this range of temperatures. As was done
for the diffusion times, fitting the data with a Vogel-Fulcher-Tammann
relation points to a divergence temperature $T_{0}\simeq230/235\:\mathrm{K}$.
The correct qualitative behavior of the results is well reproduced,
together with accurate slopes as a functionn of temperature, with
anions diffusing slower than cations, despite their smaller size.
The higher conformational flexibility of the cation, due to the presence
of alkyl chains, is generally acknowledged as the main explanation
for this behavior. Nonetheless, computed results are one order of
magnitude far from the experimental data available at the lowest temperatures
\cite{tokuda_jpcb_2004}. This discrepancy may be due to the intrinsic
deficiency of the non polarizable force-field to reproduce the real
dynamics of the systems \cite{yan_jcpb_2004}. On the other hand,
the presence of even a small fraction of impurities in the experimental
samples can be responsible for a significant change in the reported
results \cite{rollet_jpcb_2007}. 

It is important to note that original results on the LHW2004 force
field \cite{liu_jpcb_2004} showed an impressive agreement with experiments.
This apparent contradiction can be explained by the very short simulation
times used to extrapolate the diffusion coefficients in Ref. \cite{liu_jpcb_2004}.
Especially for results at the lowest temperature, according to the
considerations above and the results in Figure \ref{fig: msd_times},
a sub-linear non-diffusive behavior of the MSD should be inferred
for all simulation times reported in Ref. \cite{liu_jpcb_2004}. Longer
simulations on the same systems at $350\:\mathrm{K}$ appeared in
the literature \cite{tsuzuki_jpcb_2009} and showed a remarkable agreement
with the present results, thus validating our conclusions.

\begin{figure}
\includegraphics[width=1\textwidth]{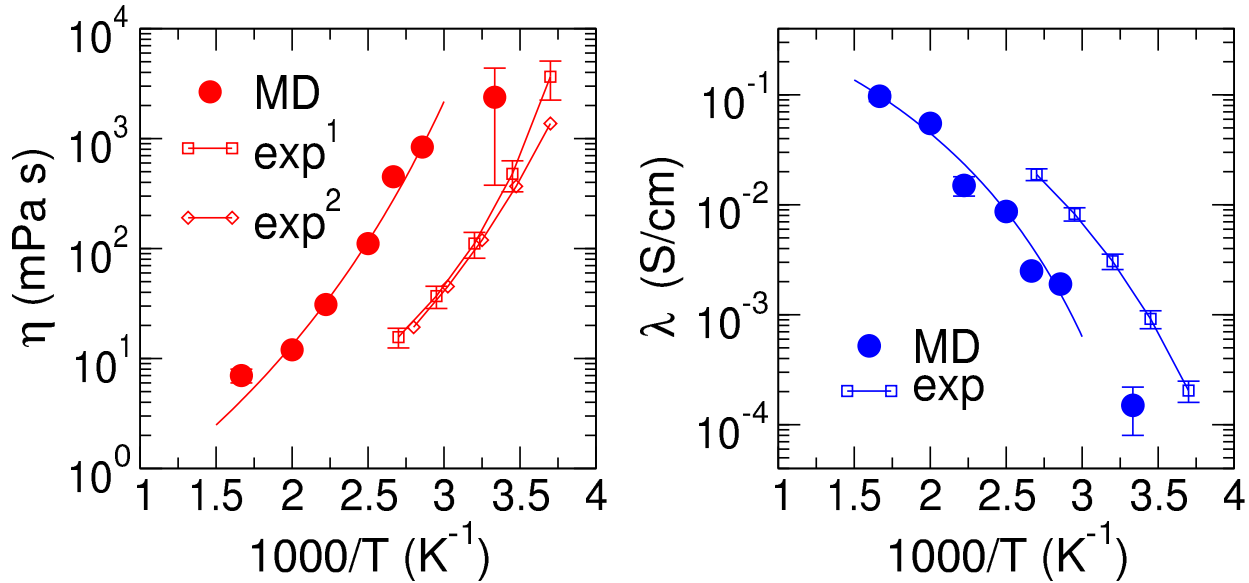}\caption{\label{fig: viscosity_and_conductivity}Bulk viscosities (left) and
ionic conductivities (right) of BMIM-PF6, as computed by classical
NVT molecular dynamics simulations on a periodic cell of 128 ion pairs
described by the LHW2004 force field. Experimental results from Ref.\cite{tokuda_jpcb_2004}(exp$^{1}$,
empty squares) and Ref.\cite{jin_jpcb_2008} (exp$^{2}$, empty diamonds)
are also reported for comparison.}
\end{figure}

Viscosities and ionic conductivities of the system are reported in
Figure \ref{fig: viscosity_and_conductivity} and summarized in Table
\ref{tab: pf6_temperature}. 
 As for the case of diffusion coefficients, also viscosities and conductivities
show significant deviations from experimental results. In both cases,
results one order of magnitude lower than the values reported in the
literature were obtained, even if their trends with respect to temperature
are also well reproduced. It should be stressed that the uncertainties
in the computed values for these quantities are higher than the average
errors in computing diffusion coefficients: this is due to the fact
that viscosity and conductivity are global properties of the system,
while diffusion coefficients calculations benefit from ensemble averages
over all different particles in the system.

\subsubsection{Size effects}

As shown by the radial distribution functions (Figure \ref{fig: rdf_detailed_128ionpairs}),
Coulomb-induced ordering of the systems is significant for distances
comparable and larger than the 128 ion pairs simulation cell. Thus,
particular care should be taken in evaluating size effects on the
structure and on the transport properties of the liquid.
\begin{figure}
\includegraphics[width=0.7\textwidth]{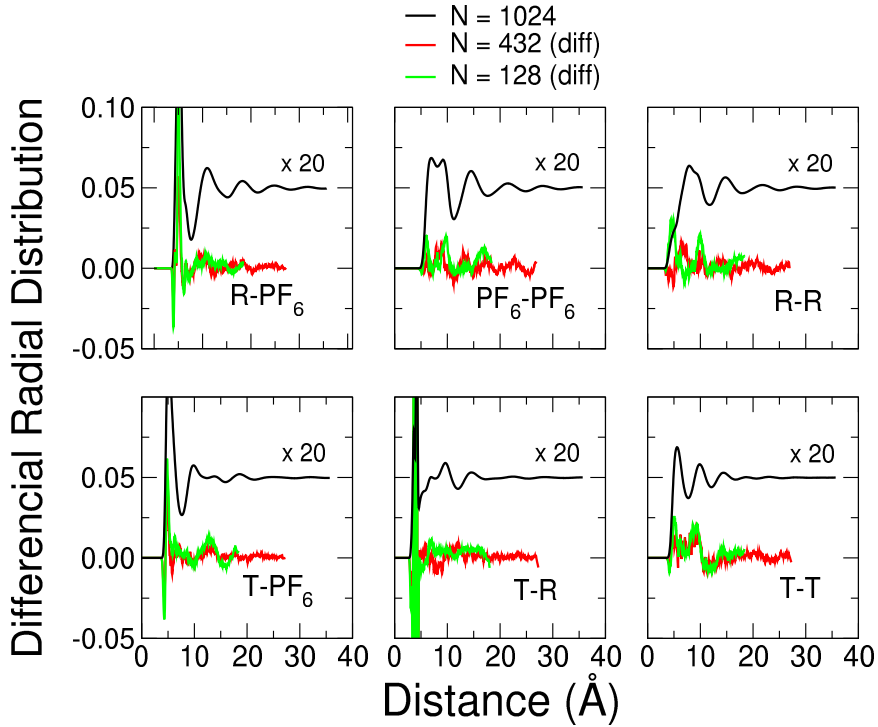}\caption{\label{fig: rdf_size}Radial distribution function of the BMIM-PF$_{6}$
RTIL, as computed by classical NVT molecular dynamics simulations
at 400 K on periodic cell of 128, 432 and 1024 ion pairs using the
LHW2004 force field. The BMIM cation is divided in two distinct subregions,
corresponding to the charged aromatic ring (R) and the long alkylic
tail (T) (see Figure \ref{fig: rdf_detailed_128ionpairs}). While
correlations involving non charged tails disappear for distances as
low as 20 angstroms, the radial distribution functions of the larger
system (black curves, scaled by a factor of twenty) show that charge
induced ordering is negligible only for distances beyond 35 Å. The
difference between the radial distribution function computed on the
larger system and the ones obtained with smaller cells are reported
in green (red) for the 128 (432) ion pairs systems. Despite the fact
that, in the smaller systems, correlation lengths are larger than
half of the cell sizes, no significant changes in the radial distribution
functions are found. }
\end{figure}

In Figure \ref{fig: rdf_size} the radial distribution functions as
a function of the number of ion pairs (128, 432, 1024) are reported
at $400\:\mathrm{K}$. Correlations due to the electrostatic interactions
in the liquid decay completely at half the cell size only for the
larger systems, composed by 1024 ion pairs. Nonetheless, the local
structure of the liquid is well reproduced in the smaller systems,
the differences in the computed radial distribution functions being
generally less than 2\%.
\begin{table}
\begin{tabular}{|c|c|c|c|c|c|c|}
\hline 
Systems & $T\left(\mathrm{K}\right)$ & $\rho$$\left(\mathrm{g}/\mathrm{c}\mathrm{m}^{3}\right)$ & $D^{+}\left(\mathrm{m}^{2}/\mathrm{s}\right)$ & $D^{-}\left(\mathrm{m}^{2}/\mathrm{s}\right)$ & $\eta\left(\mathrm{mPa\, s}\right)$ & $\lambda\left(\mathrm{S/cm}\right)$\tabularnewline
\hline 
\hline 
$N=128$ & $300$ & $1.36$ & $4.8\pm1.0\cdot10^{-13}$ & $2.6\pm0.5\cdot10^{-13}$ & $2400\pm2000$ & $1.5\pm0.7\cdot10^{-4}$\tabularnewline
\hline 
$N=432$ & $300$ & $1.36$ & $1.3\cdot10^{-13}$ & $5.0\cdot10^{-14}$ & $10000$ & $2.8\cdot10^{-5}$\tabularnewline
\hline 
$N=1024$ & $300$ & $1.37$ & $4.4\cdot10^{-13}$ & $4.0\cdot10^{-13}$ & $1700$ & $1.5\cdot10^{-4}$\tabularnewline
\hline 
$N=128$ & $400$ & $1.29$ & $3.1\pm0.1\cdot10^{-11}$ & $2.0\pm0.3\cdot10^{-11}$ & $111\pm4$ & $8.7\pm0.6\cdot10^{-3}$\tabularnewline
\hline 
$N=432$ & $400$ & $1.295$ & $2.4\cdot10^{-11}$ & $1.5\cdot10^{-11}$ & $150$ & $4.2\cdot10^{-3}$\tabularnewline
\hline 
$N=1024$ & $400$ & $1.297$ & $3.0\cdot10^{-11}$ & $2.0\cdot10^{-11}$ & $135$ & $4.7\cdot10^{-3}$\tabularnewline
\hline 
$N=128$ & $500$ & $1.22$ & $2.5\pm0.2\cdot10^{-10}$ & $1.9\pm0.1\cdot10^{-10}$ & $12\pm1$ & $4.8\pm1\cdot10^{-2}$\tabularnewline
\hline 
$N=432$ & $500$ & $1.222$ & $2.5\cdot10^{-10}$ & $1.9\cdot10^{-10}$ & $73$ & $2.4\cdot10^{-2}$\tabularnewline
\hline 
$N=1024$ & $500$ & $1.223$ & $2.9\cdot10^{-10}$ & $2.3\cdot10^{-10}$ & $37$ & $4.2\cdot10^{-2}$\tabularnewline
\hline 
\end{tabular}\caption{\label{tab: pf6_size}Densities $\rho$, cation diffusion coefficients
$D^{+}$, anion diffusion coefficients $D^{-}$, viscosities $\eta$,
and conductivities $\lambda$ of the BMIM-PF6 RTIL as a function of
temperature and system size, as computed by classical NVT molecular
dynamics simulations using the LHW2004 force field.}
\end{table}

In agreement with what was found for the distribution functions, results
in Table \ref{tab: pf6_size} show that system size affects only marginally
the computed densities. On the other hand, transport properties tend
to show more pronounced changes in going from the smallest to the
largest systems. At the lowest temperature considered (300 K), these
differences can be attributed to the very slow non ergodic dynamics
of the system and to the lack of a true diffusive regime. At the highest
temperatures, simulations seems to identify the intermediate size
system (N=432) as the least mobile one. Nonetheless, size effects
have the same order of magnitude as the uncertainty in the computed
results and values obtained already on the smaller systems give reasonable
estimates of transport properties.

\subsubsection{Effect of the force field}

In order to characterize the effect of the computational parameters
on thecalculated transport properties of the system, a comparison
between different available force fields was performed. In addition
to the results discussed in the previous sections, simulations with
the interaction potentials developed by Lopes et al. (Ref. \cite{lopes_jpcb_2004a,lopes_jpcb_2004c},
in the following abbreviated in CLDP2004), and by Morrow and Maginn
(Ref. \cite{morrow_jpcb_2002}, in the following abbreviated in MM2002),
were performed. The differences between these force fields lie mostly
in the magnitude of the electrostatic charges and in the description
of the four-body bonded interaction (dihedral angles) between the
atoms in the ring and the alkyl chains. 

In addition, a few test simulations on the CLDP2004 force field have
been performed by changing the descriptions of the intramolecular
bonds. A set of simulations with the harmonic constants of bonds and
angles reduced by half is reported, together with results obtained
by constraining all the bond lengths of the system to their equilibrium
values. 
\begin{figure}
\includegraphics[width=0.5\textwidth]{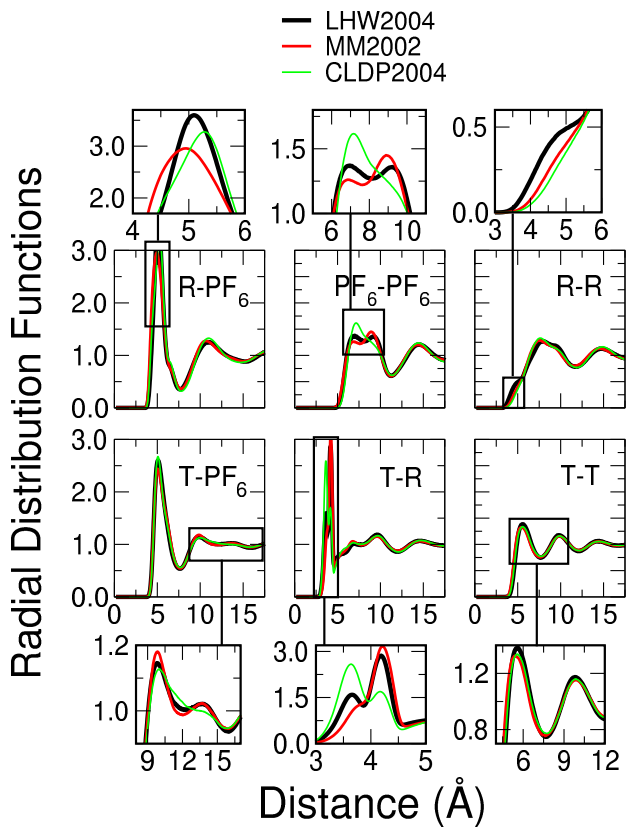}\caption{\label{fig: rdf_ff}Radial distribution function of the BMIM-PF$_{6}$
RTIL, as computed by classical NVT molecular dynamics simulations
at 400 K on periodic cell of 128 ion pairs with different force fields.
The BMIM cation is partitioned in two distinct subregions, corresponding
to the charged aromatic ring (R) and the long alkylic tail (T) (see
Figure \ref{fig: rdf_detailed_128ionpairs}). For each plot, the regions
where the radial distribution functions present the largest dependence
on the interaction potential have been highlighted in the insets.
While most of the radial distributions show only minor differences
due to the choice of the force field, the anion-anion distribution
function and the intramolecular ring-tail coordination show marked
changes in the different simulations. Moreover, only one of the considered
force fields shows the appearance of a shoulder at short distances
in the ring-ring distribution function.}
\end{figure}

For all the reported force fields, computed densities are close to
the experimental results. On the contrary, radial distribution functions
show a marked dependence on the adopted force-field (See Figure \ref{fig: rdf_ff}).
The strongest effect of the force field is on the intramolecular ring-tail
and the intermolecular anion-anion distribution functions. In the
more flexible molecules, the alkyl chain is able to come in closer
contact with the ring, thus creating some steric hindrance that interferes
with the cation-anion interactions. This is reflected in the decrease
of the second anion-anion peak, together with the slight increase
of the mean cation-anion separation. 

\begin{table}
\begin{tabular}{|c|c|c|c|c|c|c|}
\hline 
Systems & $T\left(\mathrm{K}\right)$ & $\rho$$\left(\mathrm{g}/\mathrm{c}\mathrm{m}^{3}\right)$ & $D^{+}\left(\mathrm{m}^{2}/\mathrm{s}\right)$ & $D^{-}\left(\mathrm{m}^{2}/\mathrm{s}\right)$ & $\eta\left(\mathrm{mPa\, s}\right)$ & $\lambda\left(\mathrm{S/cm}\right)$\tabularnewline
\hline 
\hline 
LHW2004 & $300$ & $1.36$ & $4.8\pm1.0\cdot10^{-13}$ & $2.6\pm0.5\cdot10^{-13}$ & $2400\pm2000$ & $1.5\pm0.7\cdot10^{-4}$\tabularnewline
\hline 
MM2002 & $300$ & $1.353$ & $3.2\cdot10^{-12}$ & $2.1\cdot10^{-12}$ & $2200$ & $4.1\cdot10^{-4}$\tabularnewline
\hline 
CLDP2004 & $300$ & $1.357$ & $5.0\cdot10^{-13}$ & $2.1\cdot10^{-13}$ & $7600$ & $2.3\cdot10^{-5}$\tabularnewline
\hline 
LHW2004 & $350$ & $1.33$ & $4.0\pm1.0\cdot10^{-12}$ & $2.0\pm0.8\cdot10^{-12}$ & $940\pm400$ & $1.9\pm0.8\cdot10^{-3}$\tabularnewline
\hline 
MM2002 & $350$ & $1.305$ & $1.3\cdot10^{-11}$ & $9.4\cdot10^{-12}$ & $77$ & $3.9\cdot10^{-3}$\tabularnewline
\hline 
CLDP2004 & $350$ & $1.318$ & $4.2\cdot10^{-12}$ & $2.3\cdot10^{-12}$ & $903$ & $7.6\cdot10^{-4}$\tabularnewline
\hline 
LHW2004 & $400$ & $1.29$ & $3.1\pm0.1\cdot10^{-11}$ & $2.0\pm0.3\cdot10^{-11}$ & $111\pm4$ & $8.7\pm0.6\cdot10^{-3}$\tabularnewline
\hline 
MM2002 & $400$ & $1.268$ & $8.0\cdot10^{-11}$ & $5.7\cdot10^{-11}$ & $26$ & $9.8\cdot10^{-3}$\tabularnewline
\hline 
CLDP2004 & $400$ & $1.275$ & $2.5\cdot10^{-11}$ & $1.7\cdot10^{-11}$ & $196$ & $7.4\cdot10^{-3}$\tabularnewline
\hline 
LHW2004 & $500$ & $1.22$ & $2.5\pm0.2\cdot10^{-10}$ & $1.9\pm0.1\cdot10^{-10}$ & $12\pm1$ & $4.8\pm1.0\cdot10^{-2}$\tabularnewline
\hline 
MM2002 & $500$ & $1.185$ & $4.7\cdot10^{-10}$ & $4.0\cdot10^{-10}$ & $10$ & $6.1\cdot10^{-2}$\tabularnewline
\hline 
CLDP2004 & $500$ & $1.200$ & $2.4\cdot10^{-10}$ & $2.1\cdot10^{-10}$ & $12$ & $2.7\cdot10^{-2}$\tabularnewline
\hline 
\end{tabular}\caption{\label{tab: pf6_forcefield}Densities $\rho$, cation diffusion coefficients
$D^{+}$, anion diffusion coefficients $D^{-}$, viscosities $\eta$,
and conductivities $\lambda$ of the BMIM-PF$_{6}$ RTIL as a function
of temperature and interaction potentials, as computed by classical
NVT molecular dynamics simulations on a system composed by 128 ion
pairs.}
\end{table}
\begin{figure}
\includegraphics[width=0.5\textwidth]{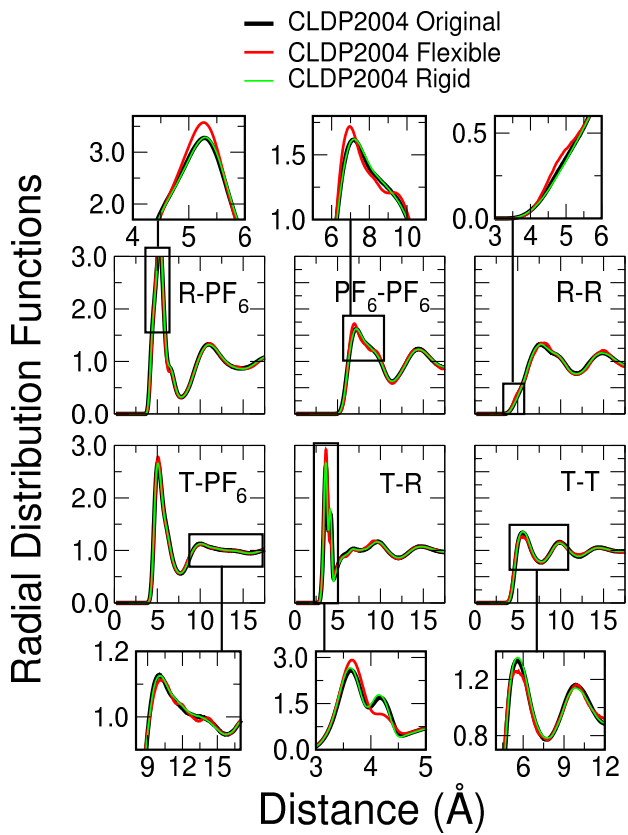}\caption{\label{fig: rdf_ff_cldp}Radial distribution function of the BMIM-PF$_{6}$
RTIL, as computed by classical NVT molecular dynamics simulations
at 400 K on periodic cell of 128 ion pairs described by the CLDP2004
force field. For each plot, the regions where the radial distribution
functions present the largest dependence on the interaction potential
have been highlighted in the insets. Different choices of the strength
of bonds and angles have been adopted, corresponding to the one reported
in the original paper \cite{lopes_jpcb_2004a,lopes_jpcb_2004c} (CLDP2004
Original), to an interaction potential with all the bonds constrained
to their equilibrium positions (CLDP2004 Rigid) and to an interaction
potential with all the harmonic constants of bonds and angles reduced
by half (CLDP Flexible). While the description of the intramolecular
bonds appear to affect only marginally the structure of the liquid,
increasing the flexibility of the harmonic three body interactions
(angles) has a marked effect on the computed distribution functions.
In particular, the resulting liquid show a higher correlation between
the charged residues at the expenses of the short-ranged tail tail
distribution.}
\end{figure}
\begin{table}
\begin{tabular}{|c|c|c|c|c|c|c|}
\hline 
Systems & $T\left(\mathrm{K}\right)$ & $\rho$$\left(\mathrm{g}/\mathrm{c}\mathrm{m}^{3}\right)$ & $D^{+}\left(\mathrm{m}^{2}/\mathrm{s}\right)$ & $D^{-}\left(\mathrm{m}^{2}/\mathrm{s}\right)$ & $\eta\left(\mathrm{mPa\, s}\right)$ & $\lambda\left(\mathrm{S/cm}\right)$\tabularnewline
\hline 
\hline 
CLDP2004 & $300$ & $1.357$ & $5.0\cdot10^{-13}$ & $2.1\cdot10^{-13}$ & $7600$ & $2.3\cdot10^{-5}$\tabularnewline
\hline 
CLDP2004 Rigid & $300$ & $1.358$ & $2.8\cdot10^{-13}$ & $2.3\cdot10^{-13}$ & $8600$ & $1.2\cdot10^{-4}$\tabularnewline
\hline 
CLDP2004 Flexible & $300$ & $1.346$ & $7.0\cdot10^{-13}$ & $6.0\cdot10^{-13}$ & $600$ & $1.4\cdot10^{-4}$\tabularnewline
\hline 
CLDP2004 & $350$ & $1.318$ & $4.2\cdot10^{-12}$ & $2.3\cdot10^{-12}$ & $903$ & $7.6\cdot10^{-4}$\tabularnewline
\hline 
CLDP2004 Rigid & $350$ & $1.318$ & $3.7\cdot10^{-12}$ & $1.9\cdot10^{-12}$ & $474$ & $7.5\cdot10^{-4}$\tabularnewline
\hline 
CLDP2004 Flexible & $350$ & $1.314$ & $5.1\cdot10^{-12}$ & $2.0\cdot10^{-12}$ & $209$ & $1.6\cdot10^{-3}$\tabularnewline
\hline 
CLDP2004 & $400$ & $1.275$ & $2.5\cdot10^{-11}$ & $1.7\cdot10^{-11}$ & $196$ & $7.4\cdot10^{-3}$\tabularnewline
\hline 
CLDP2004 Rigid & $400$ & $1.280$ & $3.3\cdot10^{-11}$ & $2.0\cdot10^{-11}$ & $46$ & $6.5\cdot10^{-3}$\tabularnewline
\hline 
CLDP2004 Flexible & $400$ & $1.274$ & $3.5\cdot10^{-11}$ & $2.3\cdot10^{-11}$ & $69$ & $9.2\cdot10^{-3}$\tabularnewline
\hline 
CLDP2004 & $500$ & $1.200$ & $2.4\cdot10^{-10}$ & $2.1\cdot10^{-10}$ & $12$ & $2.7\cdot10^{-2}$\tabularnewline
\hline 
CLDP2004 Rigid & $500$ & $1.203$ & $2.6\cdot10^{-10}$ & $2.0\cdot10^{-10}$ & $28$ & $2.7\cdot10^{-2}$\tabularnewline
\hline 
CLDP2004 Flexible & $500$ & $1.196$ & $2.5\cdot10^{-10}$ & $2.0\cdot10^{-10}$ & $22$ & $5.0\cdot10^{-2}$\tabularnewline
\hline 
\end{tabular}\caption{\label{tab: pf6_forcefield_cldp}Densities $\rho$, cation diffusion
coefficients $D^{+}$, anion diffusion coefficients $D^{-}$, viscosities
$\eta$, and conductivities $\lambda$ of the BMIM-PF$_{6}$ RTIL
as a function of temperature, as computed by classical NVT molecular
dynamics simulations on a system composed by 128 ion pairs and described
with the CLDP2004 interaction potential. Different choices of the
strength of bonds and angles have been adopted, corresponding to:
a) the one reported in the original paper \cite{lopes_jpcb_2004a,lopes_jpcb_2004c};
b) an interaction potential with all the bonds constrained to their
equilibrium positions (CLDP2004 Rigid); c) an interaction potential
with all the harmonic constants of bonds and angles reduced by half
(CLDP2004 Flexible). }
\end{table}
The effect of constraining all the bonds in the molecule is negligible
compared to the other parameters in the interaction potentials. On
the contrary, relaxing the force constants of bond and angles is responsible
for noticeable changes in the local structures of the liquid, increasing
the overall correlation and allowing the charged residues to come
in closer contact, at the expense of the non-polar portions of the
cations. 

Transport properties reflect what was found on the structure of the
liquid. In particular, the MM2002 potential shows significantly higher
mobility with respect to the other force fields considered. This is
probably related to the reduced flexibility of the ring-tail dihedral
angle of the cation, that favors an unfolded configuration for the
alkyl chain and broadens the cation-anion correlation function. Despite
the differences between the LHW2004 and the CLDP2004 force fields,
diffusion coefficients for the two cases are in close agreement and
almost one order of magnitude larger than the experiments. Viscosities
and ionic conductivities show a less clear trend with respect to the
force field, probably due to the uncertainty in the computed results
due to a lack of a proper ensemble averages. Nonetheless, the qualitative
picture is consistent with the results of the diffusion coefficients,
with MM2002 showing lower viscosities and higher conductivities at
all temperatures with respect to the other two force fields. Artificially
reducing or increasing the rigidity of the bonds and angles of the
CLDP2004 force field has a limited impact on transport properties,
the overall effect being within the numerical accuracy of the reported
quantities. Contrary to what was found above for the different classes
of force-fields, increasing the overall flexibility of the cation
seems to increase the overall diffusivities in the system.

\subsection{Effect of the anion}

Three of the most studied anions in RTILs have been considered, namely
the PF$_{6}^{-}$, BF$_{4}^{-}$ and Tf$_{2}$N$^{-}$ anions. In
order to treat all the systems on similar footings, force fields developed
by the same group of authors (Lopes et al., Refs. \cite{lopes_jpcb_2004a,lopes_jpcb_2004c}
and \cite{lopes_jpcb_2004b}) were chosen for all the ions. 
\begin{figure}
\includegraphics[width=0.5\textwidth]{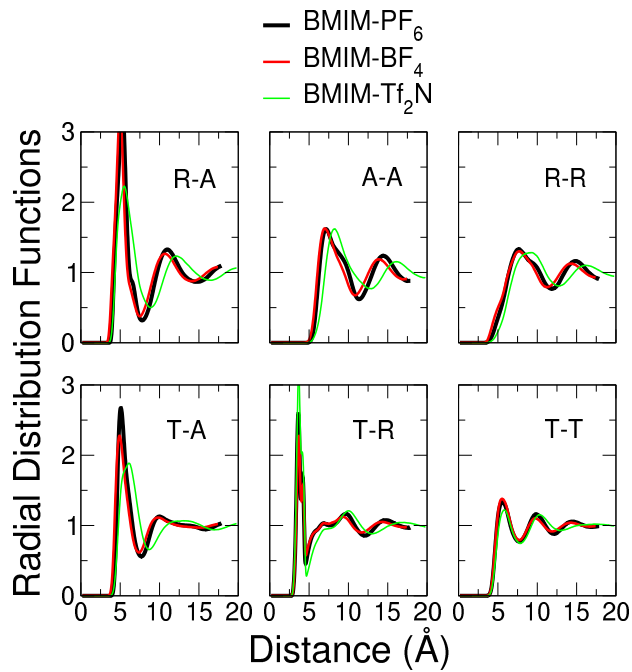}\caption{\label{fig: rdf_anion}Radial distribution function of BMIM-based
RTILs for three different anions as computed by classical NVT molecular
dynamics simulations at 400 K on periodic cell of 128 ion pairs described
by the CLDP2004 force field. The three anions considered (A) are PF$_{6}^{-}$
(thick black line), BF$_{4}^{-}$ (red line) and Tf$_{2}$N$^{-}$(thin
green line). The two inorganic anions show very similar distribution
functions, with almost identical first peak positions. This reflects
the similar interactions in the two systems between the fluorine atoms
of the anions and the cation. On the contrary, the larger and more
asymmetric Tf$_{2}$N$^{-}$ anion presents correlations between charged
residues that are broader and shifted towards larger distances, while
the short-ranged tail-tail interactions in the cation are only slightly
affected. }
\end{figure}
In examining the radial distribution functions, we can see that the
two inorganic anions show very similar behavior, with almost identical
first peak positions. This reflects the comparable size of the two
anions and the similar interactions in the two systems between the
fluorine atoms of the anions and the cation. On the contrary, the
larger and more asymmetric Tf$_{2}$N$^{-}$ anion presents correlations
between charged residues that are broader and shifted towards larger
distances. As expected, the cation-cation distribution functions are
the least affected by the choice of anion, in particular in the short-ranged
tail-tail interactions. 
\begin{table}
\begin{tabular}{|c|c|c|c|c|c|c|}
\hline 
Systems & $T\left(\mathrm{K}\right)$ & $\rho$$\left(\mathrm{g}/\mathrm{c}\mathrm{m}^{3}\right)$ & $D^{+}\left(\mathrm{m}^{2}/\mathrm{s}\right)$ & $D^{-}\left(\mathrm{m}^{2}/\mathrm{s}\right)$ & $\eta\left(\mathrm{mPa\, s}\right)$ & $\lambda\left(\mathrm{S/cm}\right)$\tabularnewline
\hline 
\hline 
PF$_{6}^{-}$ & $300$ & $1.36$ & $5.0\cdot10^{-13}$ & $2.1\cdot10^{-13}$ & $7600$ & $2.3\cdot10^{-5}$\tabularnewline
\hline 
BF$_{4}^{-}$ & $300$ & $1.28$ & $1.3\cdot10^{-12}$ & $9.2\cdot10^{-13}$ & $414$ & $6.1\cdot10^{-4}$\tabularnewline
\hline 
Tf$_{2}$N$^{-}$ & $300$ & $1.50$ & $2.5\cdot10^{-12}$ & $1.6\cdot10^{-12}$ & $158$ & $1.8\cdot10^{-3}$\tabularnewline
\hline 
PF$_{6}^{-}$ & $350$ & $1.32$ & $4.2\cdot10^{-12}$ & $2.3\cdot10^{-12}$ & $903$ & $7.6\cdot10^{-4}$\tabularnewline
\hline 
BF$_{4}^{-}$ & $350$ & $1.24$ & $1.5\cdot10^{-11}$ & $1.4\cdot10^{-11}$ & $157$ & $1.7\cdot10^{-3}$\tabularnewline
\hline 
Tf$_{2}$N$^{-}$ & $350$ & $1.45$ & $2.3\cdot10^{-11}$ & $1.5\cdot10^{-11}$ & $99$ & $2.2\cdot10^{-3}$\tabularnewline
\hline 
PF$_{6}^{-}$ & $400$ & $1.28$ & $2.5\cdot10^{-11}$ & $1.7\cdot10^{-11}$ & $196$ & $7.4\cdot10^{-3}$\tabularnewline
\hline 
BF$_{4}^{-}$ & $400$ & $1.20$ & $8.5\cdot10^{-11}$ & $6.8\cdot10^{-11}$ & $39$ & $8.4\cdot10^{-3}$\tabularnewline
\hline 
Tf$_{2}$N$^{-}$ & $400$ & $1.40$ & $9.7\cdot10^{-11}$ & $7.3\cdot10^{-11}$ & $40$ & $1.1\cdot10^{-2}$\tabularnewline
\hline 
PF$_{6}^{-}$ & $500$ & $1.20$ & $2.4\cdot10^{-10}$ & $2.1\cdot10^{-10}$ & $12$ & $2.7\cdot10^{-2}$\tabularnewline
\hline 
BF$_{4}^{-}$ & $500$ & $1.13$ & $4.7\cdot10^{-10}$ & $4.0\cdot10^{-10}$ & $25$ & $7.6\cdot10^{-2}$\tabularnewline
\hline 
Tf$_{2}$N$^{-}$ & $500$ & $1.32$ & $3.6\cdot10^{-10}$ & $3.2\cdot10^{-10}$ & $24$ & $4.1\cdot10^{-2}$\tabularnewline
\hline 
\end{tabular}\caption{\label{tab: anion}Densities $\rho$, cation diffusion coefficients
$D^{+}$, anion diffusion coefficients $D^{-}$, viscosities $\eta$,
and conductivities $\lambda$ of BMIM-based RTILs as a function of
temperature, as computed by classical NVT molecular dynamics simulations
on a system composed by 128 ion pairs and described with the CLDP2004
interaction potential. }
\end{table}

\begin{figure}
\includegraphics[width=0.5\textwidth]{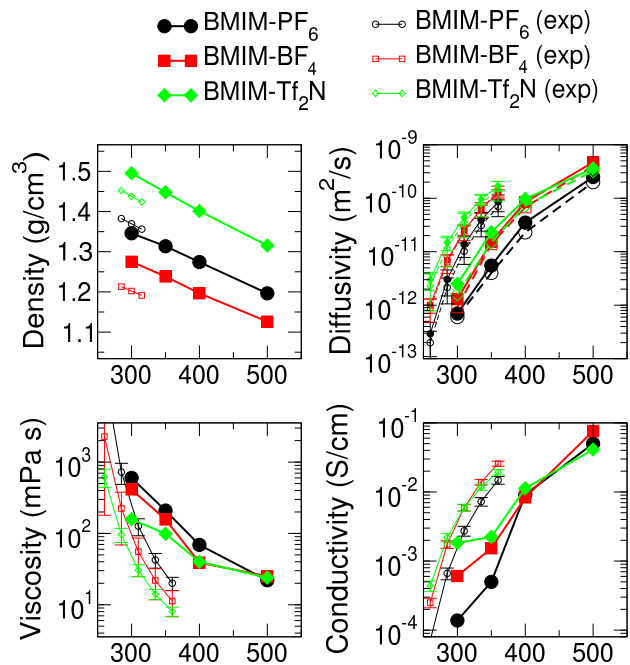}\caption{\label{fig: properties_anion}Density, diffusion coefficients for
anions (open symbols-dashed lines) and cations (filled symbols-continuous
lines), viscosity, and conductivity of BMIM-based ionic liquids for
three different anions as computed by classical NVT molecular dynamics
simulations at 400 K on a system composed of 128 ion pairs, as described
by the CLDP2004 force field. The three anions considered are PF$_{6}^{-}$
(black circles), BF$_{4}^{-}$ (red squares) and Tf$_{2}$N$^{-}$(green
diamonds). Experimental results from Ref. \cite{tokuda_jpcb_2004}
are also reported for comparison.}
\end{figure}
Densities are in good qualitative agreement with experimental results
at room temperature, but quantitative results are obtained only for
the BMIM-PF$_{6}$ system. The other two salts show deviations from
experiments much larger than the experimental uncertainty in the data.
Similarly to what was obtained for the BMIM-PF$_{6}$ RTIL, transport
properties of the liquids are within one order of magnitude from corresponding
experimental data. Nonetheless, also in this case, computed results
show a general good qualitative agreement with experiments at room
temperature. The Tf$_{2}$N salt is the one showing the highest conductivity
and diffusivity and lowest viscosity at room temperature. The PF$_{6}$
anion instead is the one giving rise to the least diffusive behavior
among the liquids in the whole temperature range considered. It is
interesting to notice that, although computed density changes are
uniform for the three liquids and linear with temperature, transport
properties show temperature dependencies strongly related to the choice
of the anion. In particular, the Tf$_{2}$N anion is the one showing
the smallest variations in the considered temperature interval. As
a consequence, already at 400 K the BF$_{4}$ liquid shows viscosities
and conductivities comparable to the organic anion.

\section{Conclusions}

In summary, a series of extensive classical MD simulations on BMIM
based ionic liquids have been reported. A careful analysis of simulation
parameters have been performed to estimate the accuracy of the computed
results. The highly viscous nature of the ionic liquids studied poses
serious problems in evaluating transport properties and was show to
require simulation times larger than 100 ns to converge below 400
K. In most cases, converged quantities could only be estimated at
room temperature, for which the systems studied are in fact trapped
in glass-like dynamics. Diffusion coefficients, viscosities and conductivities
all show good qualitative agreement and correct temperature trends
when compared with experimental results. The correct trend of the
transport properties for the different ionic liquids is recovered,
with BMIM-Tf$_{2}$N being the less viscous RTIL and BMIM-PF$_{6}$
showing the slowest dynamics. Nonetheless, results remain one order
of magnitude away from experimental data. This discrepancy could be
due to the neglect of polarizability in the interaction potentials,
or due to even small fraction of impurities in the experimental setups. 
\begin{acknowledgments}
This work was supported by E. I. du Pont de Nemours \& Co. through
the DuPont-MIT Alliance program. The authors acknowledge Shyue Ping
Ong, Gerbrand Ceder, Steve R. Lustig and William L. Holstein for useful
discussions and collaborations. 
\end{acknowledgments}
\bibliographystyle{apsrev4-1}

\begin{thebibliography}{10}%
\makeatletter
\providecommand \@ifxundefined [1]{%
 \ifx #1\undefined \expandafter \@firstoftwo
 \else \expandafter \@secondoftwo
\fi
}%
\providecommand \@ifnum [1]{%
 \ifnum #1\expandafter \@firstoftwo
 \else \expandafter \@secondoftwo
\fi
}%
\providecommand \enquote [1]{``#1''}%
\providecommand \bibnamefont  [1]{#1}%
\providecommand \bibfnamefont [1]{#1}%
\providecommand \citenamefont [1]{#1}%
\providecommand\href[0]{\@sanitize\@href}%
\providecommand\@href[1]{\endgroup\@@startlink{#1}\endgroup\@@href}%
\providecommand\@@href[1]{#1\@@endlink}%
\providecommand \@sanitize [0]{\begingroup\catcode`\&12\catcode`\#12\relax}%
\@ifxundefined \pdfoutput {\@firstoftwo}{%
 \@ifnum{\z@=\pdfoutput}{\@firstoftwo}{\@secondoftwo}%
}{%
 \providecommand\@@startlink[1]{\leavevmode}%
 \providecommand\@@endlink[0]{}%
}{%
 \providecommand\@@startlink[1]{%
  \leavevmode
  \pdfstartlink
   attr{/Border[0 0 1 ]/H/I/C[0 1 1]}%
   user{/Subtype/Link/A<</Type/Action/S/URI/URI(#1)>>}%
  \relax
 }%
 \providecommand\@@endlink[0]{\pdfendlink}%
}%
\providecommand \url  [0]{\begingroup\@sanitize \@url }%
\providecommand \@url [1]{\endgroup\@href {#1}{\urlprefix}}%
\providecommand \urlprefix [0]{URL }%
\providecommand \Eprint[0]{\href }%
\@ifxundefined \urlstyle {%
  \providecommand \doi [1]{doi:\discretionary{}{}{}#1}%
}{%
  \providecommand \doi [0]{doi:\discretionary{}{}{}\begingroup
  \urlstyle{rm}\Url }%
}%
\providecommand \doibase [0]{http://dx.doi.org/}%
\providecommand \Doi[1]{\href{\doibase#1}}%
\providecommand \bibAnnote [3]{%
  \BibitemShut{#1}%
  \begin{quotation}\noindent
    \textsc{Key:}\ #2\\\textsc{Annotation:}\ #3%
  \end{quotation}%
}%
\providecommand \bibAnnoteFile [2]{%
  \IfFileExists{#2}{\bibAnnote {#1} {#2} {\input{#2}}}{}%
}%
\providecommand \typeout [0]{\immediate \write \m@ne }%
\providecommand \selectlanguage [0]{\@gobble}%
\providecommand \bibinfo [0]{\@secondoftwo}%
\providecommand \bibfield [0]{\@secondoftwo}%
\providecommand \translation [1]{[#1]}%
\providecommand \BibitemOpen[0]{}%
\providecommand \bibitemStop [0]{}%
\providecommand \bibitemNoStop [0]{.\EOS\space}%
\providecommand \EOS [0]{\spacefactor3000\relax}%
\providecommand \BibitemShut [1]{\csname bibitem#1\endcsname}%
\bibitem{plechkova_chemsocrev_2008}%
  \BibitemOpen
  \bibfield{author}{%
  \bibinfo {author} {\bibfnamefont{N.~V.}\ \bibnamefont{Plechkova}}\ and\
  \bibinfo {author} {\bibfnamefont{K.~R.}\ \bibnamefont{Seddon}},\ }%
  \bibfield{journal}{%
  \bibinfo {journal} {Chemical Society Reviews}\ }%
  \textbf{\bibinfo {volume} {37}},\ \bibinfo {pages} {123} (\bibinfo {year}
  {2008})%
  \bibAnnoteFile{NoStop}{plechkova_chemsocrev_2008}%
\bibitem{walden_bais_1914}%
  \BibitemOpen
  \bibfield{author}{%
  \bibinfo {author} {\bibfnamefont{P.}~\bibnamefont{Walden}},\ }%
  \bibfield{journal}{%
  \bibinfo {journal} {Bulletin de l'Academie Imperiale des Sciences de
  St.Petersbourg}\ }%
  \textbf{\bibinfo {volume} {8}},\ \bibinfo {pages} {405} (\bibinfo {year}
  {1914})%
  \bibAnnoteFile{NoStop}{walden_bais_1914}%
\bibitem{wilkes_fsrl_1982}%
  \BibitemOpen
  \bibfield{author}{%
  \bibinfo {author} {\bibfnamefont{J.~S.}\ \bibnamefont{Wilkes}}\ and\ \bibinfo
  {author} {\bibfnamefont{C.~L.}\ \bibnamefont{Hussey}},\ }%
  \bibfield{journal}{%
  \bibinfo {journal} {Frank J. Seiler Research Laboratory Technical Report}}%
   (\bibinfo {year} {1982})%
  \bibAnnoteFile{NoStop}{wilkes_fsrl_1982}%
\bibitem{wilkes_inorgchem_1982}%
  \BibitemOpen
  \bibfield{author}{%
  \bibinfo {author} {\bibfnamefont{J.~S.}\ \bibnamefont{Wilkes}}, \bibinfo
  {author} {\bibfnamefont{J.~A.}\ \bibnamefont{Levisky}}, \bibinfo {author}
  {\bibfnamefont{R.~A.}\ \bibnamefont{Wilson}},\ and\ \bibinfo {author}
  {\bibfnamefont{C.~L.}\ \bibnamefont{Hussey}},\ }%
  \bibfield{journal}{%
  \bibinfo {journal} {Inorganic Chemistry}\ }%
  \textbf{\bibinfo {volume} {21}},\ \bibinfo {pages} {1263} (\bibinfo {year}
  {1982})%
  \bibAnnoteFile{NoStop}{wilkes_inorgchem_1982}%
\bibitem{hussey_pac_1988}%
  \BibitemOpen
  \bibfield{author}{%
  \bibinfo {author} {\bibfnamefont{C.~L.}\ \bibnamefont{Hussey}},\ }%
  \bibfield{journal}{%
  \bibinfo {journal} {Pure and Applied Chemistry}\ }%
  \textbf{\bibinfo {volume} {60}},\ \bibinfo {pages} {1763} (\bibinfo {year}
  {1988})%
  \bibAnnoteFile{NoStop}{hussey_pac_1988}%
\bibitem{abdulsada_chemcomm_1987}%
  \BibitemOpen
  \bibfield{author}{%
  \bibinfo {author} {\bibfnamefont{A.~K.}\ \bibnamefont{Abdulsada}}, \bibinfo
  {author} {\bibfnamefont{A.~G.}\ \bibnamefont{Avent}}, \bibinfo {author}
  {\bibfnamefont{M.~J.}\ \bibnamefont{Parkington}}, \bibinfo {author}
  {\bibfnamefont{T.~A.}\ \bibnamefont{Ryan}}, \bibinfo {author}
  {\bibfnamefont{K.~R.}\ \bibnamefont{Seddon}},\ and\ \bibinfo {author}
  {\bibfnamefont{T.}~\bibnamefont{Welton}},\ }%
  \bibfield{journal}{%
  \bibinfo {journal} {Journal of the Chemical Society-Chemical
  Communications},\ \bibinfo {pages} {1643}}%
   (\bibinfo {year} {1987})%
  \bibAnnoteFile{NoStop}{abdulsada_chemcomm_1987}%
\bibitem{abdulsada_dalton_1993}%
  \BibitemOpen
  \bibfield{author}{%
  \bibinfo {author} {\bibfnamefont{A.~K.}\ \bibnamefont{Abdulsada}}, \bibinfo
  {author} {\bibfnamefont{A.~G.}\ \bibnamefont{Avent}}, \bibinfo {author}
  {\bibfnamefont{M.~J.}\ \bibnamefont{Parkington}}, \bibinfo {author}
  {\bibfnamefont{T.~A.}\ \bibnamefont{Ryan}}, \bibinfo {author}
  {\bibfnamefont{K.~R.}\ \bibnamefont{Seddon}},\ and\ \bibinfo {author}
  {\bibfnamefont{T.}~\bibnamefont{Welton}},\ }%
  \bibfield{journal}{%
  \bibinfo {journal} {Journal of the Chemical Society-Dalton Transactions},\
  \bibinfo {pages} {3283}}%
   (\bibinfo {year} {1993})%
  \bibAnnoteFile{NoStop}{abdulsada_dalton_1993}%
\bibitem{wilkes_chemcomm_1992}%
  \BibitemOpen
  \bibfield{author}{%
  \bibinfo {author} {\bibfnamefont{J.~S.}\ \bibnamefont{Wilkes}}\ and\ \bibinfo
  {author} {\bibfnamefont{M.~J.}\ \bibnamefont{Zaworotko}},\ }%
  \bibfield{journal}{%
  \bibinfo {journal} {Journal of the Chemical Society-Chemical
  Communications},\ \bibinfo {pages} {965}}%
   (\bibinfo {year} {1992})%
  \bibAnnoteFile{NoStop}{wilkes_chemcomm_1992}%
\bibitem{weingartner_acie_2008}%
  \BibitemOpen
  \bibfield{author}{%
  \bibinfo {author} {\bibfnamefont{H.}~\bibnamefont{Weingärtner}},\ }%
  \bibfield{journal}{%
  \bibinfo {journal} {Angewandte Chemie International Edition}\ }%
  \textbf{\bibinfo {volume} {47}},\ \bibinfo {pages} {654} (\bibinfo {year}
  {2008})%
  \bibAnnoteFile{NoStop}{weingartner_acie_2008}%
\bibitem{binnemans_chemrev_2007}%
  \BibitemOpen
  \bibfield{author}{%
  \bibinfo {author} {\bibfnamefont{K.}~\bibnamefont{Binnemans}},\ }%
  \bibfield{journal}{%
  \bibinfo {journal} {Chemical Reviews}\ }%
  \textbf{\bibinfo {volume} {107}},\ \bibinfo {pages} {2592} (\bibinfo {year}
  {2007})%
  \bibAnnoteFile{NoStop}{binnemans_chemrev_2007}%
\bibitem{dupont_chemrev_2002}%
  \BibitemOpen
  \bibfield{author}{%
  \bibinfo {author} {\bibfnamefont{J.}~\bibnamefont{Dupont}}, \bibinfo {author}
  {\bibfnamefont{R.~F.}\ \bibnamefont{de~Souza}},\ and\ \bibinfo {author}
  {\bibfnamefont{P.~A.~Z.}\ \bibnamefont{Suarez}},\ }%
  \bibfield{journal}{%
  \bibinfo {journal} {Chemical Reviews}\ }%
  \textbf{\bibinfo {volume} {102}},\ \bibinfo {pages} {3667} (\bibinfo {year}
  {2002})%
  \bibAnnoteFile{NoStop}{dupont_chemrev_2002}%
\bibitem{greaves_chemrev_2008}%
  \BibitemOpen
  \bibfield{author}{%
  \bibinfo {author} {\bibfnamefont{T.~L.}\ \bibnamefont{Greaves}}\ and\
  \bibinfo {author} {\bibfnamefont{C.~J.}\ \bibnamefont{Drummond}},\ }%
  \bibfield{journal}{%
  \bibinfo {journal} {Chemical Reviews}\ }%
  \textbf{\bibinfo {volume} {108}},\ \bibinfo {pages} {206} (\bibinfo {year}
  {2008})%
  \bibAnnoteFile{NoStop}{greaves_chemrev_2008}%
\bibitem{hapiot_chemrev_2008}%
  \BibitemOpen
  \bibfield{author}{%
  \bibinfo {author} {\bibfnamefont{P.}~\bibnamefont{Hapiot}}\ and\ \bibinfo
  {author} {\bibfnamefont{C.}~\bibnamefont{Lagrost}},\ }%
  \bibfield{journal}{%
  \bibinfo {journal} {Chemical Reviews}\ }%
  \textbf{\bibinfo {volume} {108}},\ \bibinfo {pages} {2238} (\bibinfo {year}
  {2008})%
  \bibAnnoteFile{NoStop}{hapiot_chemrev_2008}%
\bibitem{vanrantwijk_chemrev_2007}%
  \BibitemOpen
  \bibfield{author}{%
  \bibinfo {author} {\bibfnamefont{F.}~\bibnamefont{van Rantwijk}}\ and\
  \bibinfo {author} {\bibfnamefont{R.~A.}\ \bibnamefont{Sheldon}},\ }%
  \bibfield{journal}{%
  \bibinfo {journal} {Chemical Reviews}\ }%
  \textbf{\bibinfo {volume} {107}},\ \bibinfo {pages} {2757} (\bibinfo {year}
  {2007})%
  \bibAnnoteFile{NoStop}{vanrantwijk_chemrev_2007}%
\bibitem{bonhote_inorgchem_1996}%
  \BibitemOpen
  \bibfield{author}{%
  \bibinfo {author} {\bibfnamefont{P.}~\bibnamefont{Bonhote}}, \bibinfo
  {author} {\bibfnamefont{A.~P.}\ \bibnamefont{Dias}}, \bibinfo {author}
  {\bibfnamefont{M.}~\bibnamefont{Armand}}, \bibinfo {author}
  {\bibfnamefont{N.}~\bibnamefont{Papageorgiou}}, \bibinfo {author}
  {\bibfnamefont{K.}~\bibnamefont{Kalyanasundaram}},\ and\ \bibinfo {author}
  {\bibfnamefont{M.}~\bibnamefont{Gratzel}},\ }%
  \bibfield{journal}{%
  \bibinfo {journal} {Inorganic Chemistry}\ }%
  \textbf{\bibinfo {volume} {35}},\ \bibinfo {pages} {1168} (\bibinfo {year}
  {1996})%
  \bibAnnoteFile{NoStop}{bonhote_inorgchem_1996}%
\bibitem{ito_naturephotonics_2008}%
  \BibitemOpen
  \bibfield{author}{%
  \bibinfo {author} {\bibfnamefont{S.}~\bibnamefont{Ito}}, \bibinfo {author}
  {\bibfnamefont{S.~M.}\ \bibnamefont{Zakeeruddin}}, \bibinfo {author}
  {\bibfnamefont{P.}~\bibnamefont{Comte}}, \bibinfo {author}
  {\bibfnamefont{P.}~\bibnamefont{Liska}}, \bibinfo {author}
  {\bibfnamefont{D.~B.}\ \bibnamefont{Kuang}},\ and\ \bibinfo {author}
  {\bibfnamefont{M.}~\bibnamefont{Gratzel}},\ }%
  \bibfield{journal}{%
  \bibinfo {journal} {Nature Photonics}\ }%
  \textbf{\bibinfo {volume} {2}},\ \bibinfo {pages} {693} (\bibinfo {year}
  {2008})%
  \bibAnnoteFile{NoStop}{ito_naturephotonics_2008}%
\bibitem{kuang_acie_2008}%
  \BibitemOpen
  \bibfield{author}{%
  \bibinfo {author} {\bibfnamefont{D.}~\bibnamefont{Kuang}}, \bibinfo {author}
  {\bibfnamefont{S.}~\bibnamefont{Uchida}}, \bibinfo {author}
  {\bibfnamefont{R.}~\bibnamefont{Humphry-Baker}}, \bibinfo {author}
  {\bibfnamefont{S.~M.}\ \bibnamefont{Zakeeruddin}},\ and\ \bibinfo {author}
  {\bibfnamefont{M.}~\bibnamefont{Gratzel}},\ }%
  \bibfield{journal}{%
  \bibinfo {journal} {Angewandte Chemie-International Edition}\ }%
  \textbf{\bibinfo {volume} {47}},\ \bibinfo {pages} {1923} (\bibinfo {year}
  {2008})%
  \bibAnnoteFile{NoStop}{kuang_acie_2008}%
\bibitem{fernicola_cpc_2007}%
  \BibitemOpen
  \bibfield{author}{%
  \bibinfo {author} {\bibfnamefont{A.}~\bibnamefont{Fernicola}}, \bibinfo
  {author} {\bibfnamefont{S.}~\bibnamefont{Panero}}, \bibinfo {author}
  {\bibfnamefont{B.}~\bibnamefont{Scrosati}}, \bibinfo {author}
  {\bibfnamefont{M.}~\bibnamefont{Tamada}},\ and\ \bibinfo {author}
  {\bibfnamefont{H.}~\bibnamefont{Ohno}},\ }%
  \bibfield{journal}{%
  \bibinfo {journal} {Chemphyschem}\ }%
  \textbf{\bibinfo {volume} {8}},\ \bibinfo {pages} {1103} (\bibinfo {year}
  {2007})%
  \bibAnnoteFile{NoStop}{fernicola_cpc_2007}%
\bibitem{fernicola_jpowersource_2007}%
  \BibitemOpen
  \bibfield{author}{%
  \bibinfo {author} {\bibfnamefont{A.}~\bibnamefont{Fernicola}}, \bibinfo
  {author} {\bibfnamefont{F.}~\bibnamefont{Croce}}, \bibinfo {author}
  {\bibfnamefont{B.}~\bibnamefont{Scrosati}}, \bibinfo {author}
  {\bibfnamefont{T.}~\bibnamefont{Watanabe}},\ and\ \bibinfo {author}
  {\bibfnamefont{H.}~\bibnamefont{Ohno}},\ }%
  \bibfield{journal}{%
  \bibinfo {journal} {Journal of Power Sources}\ }%
  \textbf{\bibinfo {volume} {174}},\ \bibinfo {pages} {342} (\bibinfo {year}
  {2007})%
  \bibAnnoteFile{NoStop}{fernicola_jpowersource_2007}%
\bibitem{seddon_pac_2000}%
  \BibitemOpen
  \bibfield{author}{%
  \bibinfo {author} {\bibfnamefont{K.~R.}\ \bibnamefont{Seddon}}, \bibinfo
  {author} {\bibfnamefont{A.}~\bibnamefont{Stark}},\ and\ \bibinfo {author}
  {\bibfnamefont{M.~J.}\ \bibnamefont{Torres}},\ }%
  \bibfield{journal}{%
  \bibinfo {journal} {Pure and Applied Chemistry}\ }%
  \textbf{\bibinfo {volume} {72}},\ \bibinfo {pages} {2275} (\bibinfo {year}
  {2000})%
  \bibAnnoteFile{NoStop}{seddon_pac_2000}%
\bibitem{rollet_jpcb_2007}%
  \BibitemOpen
  \bibfield{author}{%
  \bibinfo {author} {\bibfnamefont{A.~L.}\ \bibnamefont{Rollet}}, \bibinfo
  {author} {\bibfnamefont{P.}~\bibnamefont{Porion}}, \bibinfo {author}
  {\bibfnamefont{M.}~\bibnamefont{Vaultier}}, \bibinfo {author}
  {\bibfnamefont{I.}~\bibnamefont{Billard}}, \bibinfo {author}
  {\bibfnamefont{M.}~\bibnamefont{Deschamps}}, \bibinfo {author}
  {\bibfnamefont{C.}~\bibnamefont{Bessada}},\ and\ \bibinfo {author}
  {\bibfnamefont{L.}~\bibnamefont{Jouvensal}},\ }%
  \bibfield{journal}{%
  \bibinfo {journal} {Journal of Physical Chemistry B}\ }%
  \textbf{\bibinfo {volume} {111}},\ \bibinfo {pages} {11888} (\bibinfo {year}
  {2007})%
  \bibAnnoteFile{NoStop}{rollet_jpcb_2007}%
\bibitem{hanke_molphys_2001}%
  \BibitemOpen
  \bibfield{author}{%
  \bibinfo {author} {\bibfnamefont{C.~G.}\ \bibnamefont{Hanke}}, \bibinfo
  {author} {\bibfnamefont{S.~L.}\ \bibnamefont{Price}},\ and\ \bibinfo {author}
  {\bibfnamefont{R.~M.}\ \bibnamefont{Lynden-Bell}},\ }%
  \bibfield{journal}{%
  \bibinfo {journal} {Molecular Physics}\ }%
  \textbf{\bibinfo {volume} {99}},\ \bibinfo {pages} {801} (\bibinfo {year}
  {2001})%
  \bibAnnoteFile{NoStop}{hanke_molphys_2001}%
\bibitem{morrow_jpcb_2002}%
  \BibitemOpen
  \bibfield{author}{%
  \bibinfo {author} {\bibfnamefont{T.~I.}\ \bibnamefont{Morrow}}\ and\ \bibinfo
  {author} {\bibfnamefont{E.~J.}\ \bibnamefont{Maginn}},\ }%
  \bibfield{journal}{%
  \bibinfo {journal} {Journal of Physical Chemistry B}\ }%
  \textbf{\bibinfo {volume} {106}},\ \bibinfo {pages} {12807} (\bibinfo {year}
  {2002})%
  \bibAnnoteFile{NoStop}{morrow_jpcb_2002}%
\bibitem{shah_greenchem_2002}%
  \BibitemOpen
  \bibfield{author}{%
  \bibinfo {author} {\bibfnamefont{J.~K.}\ \bibnamefont{Shah}}, \bibinfo
  {author} {\bibfnamefont{J.~F.}\ \bibnamefont{Brennecke}},\ and\ \bibinfo
  {author} {\bibfnamefont{E.~J.}\ \bibnamefont{Maginn}},\ }%
  \bibfield{journal}{%
  \bibinfo {journal} {Green Chemistry}\ }%
  \textbf{\bibinfo {volume} {4}},\ \bibinfo {pages} {112} (\bibinfo {year}
  {2002})%
  \bibAnnoteFile{NoStop}{shah_greenchem_2002}%
\bibitem{shah_fluidphaseeq_2004}%
  \BibitemOpen
  \bibfield{author}{%
  \bibinfo {author} {\bibfnamefont{J.~K.}\ \bibnamefont{Shah}}\ and\ \bibinfo
  {author} {\bibfnamefont{E.~J.}\ \bibnamefont{Maginn}},\ }%
  \bibfield{journal}{%
  \bibinfo {journal} {Fluid Phase Equilibria}\ }%
  \textbf{\bibinfo {volume} {222}},\ \bibinfo {pages} {195} (\bibinfo {year}
  {2004})%
  \bibAnnoteFile{NoStop}{shah_fluidphaseeq_2004}%
\bibitem{margulis_jpcb_2002}%
  \BibitemOpen
  \bibfield{author}{%
  \bibinfo {author} {\bibfnamefont{C.~J.}\ \bibnamefont{Margulis}}, \bibinfo
  {author} {\bibfnamefont{H.~A.}\ \bibnamefont{Stern}},\ and\ \bibinfo {author}
  {\bibfnamefont{B.~J.}\ \bibnamefont{Berne}},\ }%
  \bibfield{journal}{%
  \bibinfo {journal} {Journal of Physical Chemistry B}\ }%
  \textbf{\bibinfo {volume} {106}},\ \bibinfo {pages} {12017} (\bibinfo {year}
  {2002})%
  \bibAnnoteFile{NoStop}{margulis_jpcb_2002}%
\bibitem{deandrade_jpcb_2002}%
  \BibitemOpen
  \bibfield{author}{%
  \bibinfo {author} {\bibfnamefont{J.}~\bibnamefont{de~Andrade}}, \bibinfo
  {author} {\bibfnamefont{E.~S.}\ \bibnamefont{Boes}},\ and\ \bibinfo {author}
  {\bibfnamefont{H.}~\bibnamefont{Stassen}},\ }%
  \bibfield{journal}{%
  \bibinfo {journal} {Journal of Physical Chemistry B}\ }%
  \textbf{\bibinfo {volume} {106}},\ \bibinfo {pages} {3546} (\bibinfo {year}
  {2002})%
  \bibAnnoteFile{NoStop}{deandrade_jpcb_2002}%
\bibitem{deandrade_jpcb_2002b}%
  \BibitemOpen
  \bibfield{author}{%
  \bibinfo {author} {\bibfnamefont{J.}~\bibnamefont{de~Andrade}}, \bibinfo
  {author} {\bibfnamefont{E.~S.}\ \bibnamefont{Boes}},\ and\ \bibinfo {author}
  {\bibfnamefont{H.}~\bibnamefont{Stassen}},\ }%
  \bibfield{journal}{%
  \bibinfo {journal} {Journal of Physical Chemistry B}\ }%
  \textbf{\bibinfo {volume} {106}},\ \bibinfo {pages} {13344} (\bibinfo {year}
  {2002})%
  \bibAnnoteFile{NoStop}{deandrade_jpcb_2002b}%
\bibitem{hunt_molsim_2006}%
  \BibitemOpen
  \bibfield{author}{%
  \bibinfo {author} {\bibfnamefont{P.~A.}\ \bibnamefont{Hunt}},\ }%
  \bibfield{journal}{%
  \bibinfo {journal} {Molecular Simulation}\ }%
  \textbf{\bibinfo {volume} {32}},\ \bibinfo {pages} {1} (\bibinfo {year}
  {2006})%
  \bibAnnoteFile{NoStop}{hunt_molsim_2006}%
\bibitem{yan_jcpb_2004}%
  \BibitemOpen
  \bibfield{author}{%
  \bibinfo {author} {\bibfnamefont{T.~Y.}\ \bibnamefont{Yan}}, \bibinfo
  {author} {\bibfnamefont{C.~J.}\ \bibnamefont{Burnham}}, \bibinfo {author}
  {\bibfnamefont{M.~G.}\ \bibnamefont{Del~Popolo}},\ and\ \bibinfo {author}
  {\bibfnamefont{G.~A.}\ \bibnamefont{Voth}},\ }%
  \bibfield{journal}{%
  \bibinfo {journal} {Journal of Physical Chemistry B}\ }%
  \textbf{\bibinfo {volume} {108}},\ \bibinfo {pages} {11877} (\bibinfo {year}
  {2004})%
  \bibAnnoteFile{NoStop}{yan_jcpb_2004}%
\bibitem{hu_pnas_2007}%
  \BibitemOpen
  \bibfield{author}{%
  \bibinfo {author} {\bibfnamefont{Z.~H.}\ \bibnamefont{Hu}}\ and\ \bibinfo
  {author} {\bibfnamefont{C.~J.}\ \bibnamefont{Margulis}},\ }%
  \bibfield{journal}{%
  \bibinfo {journal} {Proceedings of the National Academy of Sciences of the
  United States of America}\ }%
  \textbf{\bibinfo {volume} {104}},\ \bibinfo {pages} {9546} (\bibinfo {year}
  {2007})%
  \bibAnnoteFile{NoStop}{hu_pnas_2007}%
\bibitem{habasaki_jcp_2008}%
  \BibitemOpen
  \bibfield{author}{%
  \bibinfo {author} {\bibfnamefont{J.}~\bibnamefont{Habasaki}}\ and\ \bibinfo
  {author} {\bibfnamefont{K.~L.}\ \bibnamefont{Ngai}},\ }%
  \bibfield{journal}{%
  \bibinfo {journal} {Journal of Chemical Physics}\ }%
  \textbf{\bibinfo {volume} {129}},\ \bibinfo {pages} {194501} (\bibinfo {year}
  {2008})%
  \bibAnnoteFile{NoStop}{habasaki_jcp_2008}%
\bibitem{qiao_jpcb_2008}%
  \BibitemOpen
  \bibfield{author}{%
  \bibinfo {author} {\bibfnamefont{B.}~\bibnamefont{Qiao}}, \bibinfo {author}
  {\bibfnamefont{C.}~\bibnamefont{Krekeler}}, \bibinfo {author}
  {\bibfnamefont{R.}~\bibnamefont{Berger}}, \bibinfo {author}
  {\bibfnamefont{L.}~\bibnamefont{Delle~Site}},\ and\ \bibinfo {author}
  {\bibfnamefont{C.}~\bibnamefont{Holm}},\ }%
  \bibfield{journal}{%
  \bibinfo {journal} {Journal of Physical Chemistry B}\ }%
  \textbf{\bibinfo {volume} {112}},\ \bibinfo {pages} {1743} (\bibinfo {year}
  {2008})%
  \bibAnnoteFile{NoStop}{qiao_jpcb_2008}%
\bibitem{cadena_jpcb_2006b}%
  \BibitemOpen
  \bibfield{author}{%
  \bibinfo {author} {\bibfnamefont{C.}~\bibnamefont{Cadena}}, \bibinfo {author}
  {\bibfnamefont{Q.}~\bibnamefont{Zhao}}, \bibinfo {author}
  {\bibfnamefont{R.~Q.}\ \bibnamefont{Snurr}},\ and\ \bibinfo {author}
  {\bibfnamefont{E.~J.}\ \bibnamefont{Maginn}},\ }%
  \bibfield{journal}{%
  \bibinfo {journal} {Journal of Physical Chemistry B}\ }%
  \textbf{\bibinfo {volume} {110}},\ \bibinfo {pages} {2821} (\bibinfo {year}
  {2006})%
  \bibAnnoteFile{NoStop}{cadena_jpcb_2006b}%
\bibitem{reycastro_jpcb_2006a}%
  \BibitemOpen
  \bibfield{author}{%
  \bibinfo {author} {\bibfnamefont{C.}~\bibnamefont{Rey-Castro}}\ and\ \bibinfo
  {author} {\bibfnamefont{L.~F.}\ \bibnamefont{Vega}},\ }%
  \bibfield{journal}{%
  \bibinfo {journal} {Journal of Physical Chemistry B}\ }%
  \textbf{\bibinfo {volume} {110}},\ \bibinfo {pages} {14426} (\bibinfo {year}
  {2006})%
  \bibAnnoteFile{NoStop}{reycastro_jpcb_2006a}%
\bibitem{reycastro_jpcb_2006b}%
  \BibitemOpen
  \bibfield{author}{%
  \bibinfo {author} {\bibfnamefont{C.}~\bibnamefont{Rey-Castro}}\ and\ \bibinfo
  {author} {\bibfnamefont{L.~F.}\ \bibnamefont{Vega}},\ }%
  \bibfield{journal}{%
  \bibinfo {journal} {Journal of Physical Chemistry B}\ }%
  \textbf{\bibinfo {volume} {110}},\ \bibinfo {pages} {16157} (\bibinfo {year}
  {2006})%
  \bibAnnoteFile{NoStop}{reycastro_jpcb_2006b}%
\bibitem{smith_jmolgraph_1996}%
  \BibitemOpen
  \bibfield{author}{%
  \bibinfo {author} {\bibfnamefont{W.}~\bibnamefont{Smith}}\ and\ \bibinfo
  {author} {\bibfnamefont{T.~R.}\ \bibnamefont{Forester}},\ }%
  \bibfield{journal}{%
  \bibinfo {journal} {Journal of Molecular Graphics}\ }%
  \textbf{\bibinfo {volume} {14}},\ \bibinfo {pages} {136} (\bibinfo {year}
  {1996})%
  \bibAnnoteFile{NoStop}{smith_jmolgraph_1996}%
\bibitem{smith_molsim_2002}%
  \BibitemOpen
  \bibfield{author}{%
  \bibinfo {author} {\bibfnamefont{W.}~\bibnamefont{Smith}}, \bibinfo {author}
  {\bibfnamefont{C.~W.}\ \bibnamefont{Yong}},\ and\ \bibinfo {author}
  {\bibfnamefont{P.~M.}\ \bibnamefont{Rodger}},\ }%
  \bibfield{journal}{%
  \bibinfo {journal} {Molecular Simulation}\ }%
  \textbf{\bibinfo {volume} {28}},\ \bibinfo {pages} {385} (\bibinfo {year}
  {2002})%
  \bibAnnoteFile{NoStop}{smith_molsim_2002}%
\bibitem{smith_molsim_2006}%
  \BibitemOpen
  \bibfield{author}{%
  \bibinfo {author} {\bibfnamefont{W.}~\bibnamefont{Smith}},\ }%
  \bibfield{journal}{%
  \bibinfo {journal} {Molecular Simulation}\ }%
  \textbf{\bibinfo {volume} {32}},\ \bibinfo {pages} {933} (\bibinfo {year}
  {2006})%
  \bibAnnoteFile{NoStop}{smith_molsim_2006}%
\bibitem{liu_jpcb_2004}%
  \BibitemOpen
  \bibfield{author}{%
  \bibinfo {author} {\bibfnamefont{Z.~P.}\ \bibnamefont{Liu}}, \bibinfo
  {author} {\bibfnamefont{S.~P.}\ \bibnamefont{Huang}},\ and\ \bibinfo {author}
  {\bibfnamefont{W.~C.}\ \bibnamefont{Wang}},\ }%
  \bibfield{journal}{%
  \bibinfo {journal} {Journal of Physical Chemistry B}\ }%
  \textbf{\bibinfo {volume} {108}},\ \bibinfo {pages} {12978} (\bibinfo {year}
  {2004})%
  \bibAnnoteFile{NoStop}{liu_jpcb_2004}%
\bibitem{lopes_jpcb_2004a}%
  \BibitemOpen
  \bibfield{author}{%
  \bibinfo {author} {\bibfnamefont{J.~N.~C.}\ \bibnamefont{Lopes}}, \bibinfo
  {author} {\bibfnamefont{J.}~\bibnamefont{Deschamps}},\ and\ \bibinfo {author}
  {\bibfnamefont{A.~A.~H.}\ \bibnamefont{Padua}},\ }%
  \bibfield{journal}{%
  \bibinfo {journal} {Journal of Physical Chemistry B}\ }%
  \textbf{\bibinfo {volume} {108}},\ \bibinfo {pages} {2038} (\bibinfo {year}
  {2004})%
  \bibAnnoteFile{NoStop}{lopes_jpcb_2004a}%
\bibitem{lopes_jpcb_2004c}%
  \BibitemOpen
  \bibfield{author}{%
  \bibinfo {author} {\bibfnamefont{J.~N.~C.}\ \bibnamefont{Lopes}}, \bibinfo
  {author} {\bibfnamefont{J.}~\bibnamefont{Deschamps}},\ and\ \bibinfo {author}
  {\bibfnamefont{A.~A.~H.}\ \bibnamefont{Padua}},\ }%
  \bibfield{journal}{%
  \bibinfo {journal} {Journal of Physical Chemistry B}\ }%
  \textbf{\bibinfo {volume} {108}},\ \bibinfo {pages} {11250} (\bibinfo {year}
  {2004})%
  \bibAnnoteFile{NoStop}{lopes_jpcb_2004c}%
\bibitem{lopes_jpcb_2004b}%
  \BibitemOpen
  \bibfield{author}{%
  \bibinfo {author} {\bibfnamefont{J.~N.~C.}\ \bibnamefont{Lopes}}\ and\
  \bibinfo {author} {\bibfnamefont{A.~A.~H.}\ \bibnamefont{Padua}},\ }%
  \bibfield{journal}{%
  \bibinfo {journal} {Journal of Physical Chemistry B}\ }%
  \textbf{\bibinfo {volume} {108}},\ \bibinfo {pages} {16893} (\bibinfo {year}
  {2004})%
  \bibAnnoteFile{NoStop}{lopes_jpcb_2004b}%
\bibitem{berendsen_jcp_1984}%
  \BibitemOpen
  \bibfield{author}{%
  \bibinfo {author} {\bibfnamefont{H.~J.~C.}\ \bibnamefont{Berendsen}},
  \bibinfo {author} {\bibfnamefont{J.~P.~M.}\ \bibnamefont{Postma}}, \bibinfo
  {author} {\bibfnamefont{W.~F.}\ \bibnamefont{Vangunsteren}}, \bibinfo
  {author} {\bibfnamefont{A.}~\bibnamefont{Dinola}},\ and\ \bibinfo {author}
  {\bibfnamefont{J.~R.}\ \bibnamefont{Haak}},\ }%
  \bibfield{journal}{%
  \bibinfo {journal} {Journal of Chemical Physics}\ }%
  \textbf{\bibinfo {volume} {81}},\ \bibinfo {pages} {3684} (\bibinfo {year}
  {1984})%
  \bibAnnoteFile{NoStop}{berendsen_jcp_1984}%
\bibitem{nose_molphys_1984}%
  \BibitemOpen
  \bibfield{author}{%
  \bibinfo {author} {\bibfnamefont{S.}~\bibnamefont{Nose}},\ }%
  \bibfield{journal}{%
  \bibinfo {journal} {Molecular Physics}\ }%
  \textbf{\bibinfo {volume} {52}},\ \bibinfo {pages} {255} (\bibinfo {year}
  {1984})%
  \bibAnnoteFile{NoStop}{nose_molphys_1984}%
\bibitem{hoover_pra_1985}%
  \BibitemOpen
  \bibfield{author}{%
  \bibinfo {author} {\bibfnamefont{W.~G.}\ \bibnamefont{Hoover}},\ }%
  \bibfield{journal}{%
  \bibinfo {journal} {Physical Review A}\ }%
  \textbf{\bibinfo {volume} {31}},\ \bibinfo {pages} {1695} (\bibinfo {year}
  {1985})%
  \bibAnnoteFile{NoStop}{hoover_pra_1985}%
\bibitem{frenkel_book_2001}%
  \BibitemOpen
  \bibfield{author}{%
  \bibinfo {author} {\bibfnamefont{D.}~\bibnamefont{Frenkel}}\ and\ \bibinfo
  {author} {\bibfnamefont{B.}~\bibnamefont{Smit}},\ }%
  \emph{\bibinfo {title} {Understanding Molecular Simulation, From Algorithms
  to Applications}},\ \bibinfo {edition} {2nd}\ ed.,\ \bibinfo {series}
  {Computational Science, From Theory to Applications}, Vol.~\bibinfo {volume}
  {1}\ (\bibinfo {publisher} {Academic Press, Inc.},\ \bibinfo {year} {2001})%
  \bibAnnoteFile{NoStop}{frenkel_book_2001}%
\bibitem{allen_book_1989}%
  \BibitemOpen
  \bibfield{author}{%
  \bibinfo {author} {\bibfnamefont{M.}~\bibnamefont{Allen}}\ and\ \bibinfo
  {author} {\bibfnamefont{D.~J.}\ \bibnamefont{Tildesley}},\ }%
  \emph{\bibinfo {title} {Computer Simulation of Liquids}},\ \bibinfo {edition}
  {paperback}\ ed.\ (\bibinfo {publisher} {Oxford University Press, USA},\
  \bibinfo {year} {1989})%
  \bibAnnoteFile{NoStop}{allen_book_1989}%
\bibitem{tokuda_jpcb_2004}%
  \BibitemOpen
  \bibfield{author}{%
  \bibinfo {author} {\bibfnamefont{H.}~\bibnamefont{Tokuda}}, \bibinfo {author}
  {\bibfnamefont{K.}~\bibnamefont{Hayamizu}}, \bibinfo {author}
  {\bibfnamefont{K.}~\bibnamefont{Ishii}}, \bibinfo {author}
  {\bibfnamefont{M.}~\bibnamefont{Abu Bin Hasan~Susan}},\ and\ \bibinfo
  {author} {\bibfnamefont{M.}~\bibnamefont{Watanabe}},\ }%
  \bibfield{journal}{%
  \bibinfo {journal} {Journal of Physical Chemistry B}\ }%
  \textbf{\bibinfo {volume} {108}},\ \bibinfo {pages} {16593} (\bibinfo {year}
  {2004})%
  \bibAnnoteFile{NoStop}{tokuda_jpcb_2004}%
\bibitem{jin_jpcb_2008}%
  \BibitemOpen
  \bibfield{author}{%
  \bibinfo {author} {\bibfnamefont{H.}~\bibnamefont{Jin}}, \bibinfo {author}
  {\bibfnamefont{B.}~\bibnamefont{O'Hare}}, \bibinfo {author}
  {\bibfnamefont{J.}~\bibnamefont{Dong}}, \bibinfo {author}
  {\bibfnamefont{S.}~\bibnamefont{Arzhantsev}}, \bibinfo {author}
  {\bibfnamefont{G.~A.}\ \bibnamefont{Baker}}, \bibinfo {author}
  {\bibfnamefont{J.~F.}\ \bibnamefont{Wishart}}, \bibinfo {author}
  {\bibfnamefont{A.~J.}\ \bibnamefont{Benesi}},\ and\ \bibinfo {author}
  {\bibfnamefont{M.}~\bibnamefont{Maroncelli}},\ }%
  \bibfield{journal}{%
  \bibinfo {journal} {Journal of Physical Chemistry B}\ }%
  \textbf{\bibinfo {volume} {112}},\ \bibinfo {pages} {81} (\bibinfo {year}
  {2008})%
  \bibAnnoteFile{NoStop}{jin_jpcb_2008}%
\bibitem{tsuzuki_jpcb_2009}%
  \BibitemOpen
  \bibfield{author}{%
  \bibinfo {author} {\bibfnamefont{S.}~\bibnamefont{Tsuzuki}}, \bibinfo
  {author} {\bibfnamefont{W.}~\bibnamefont{Shinoda}}, \bibinfo {author}
  {\bibfnamefont{H.}~\bibnamefont{Saito}}, \bibinfo {author}
  {\bibfnamefont{M.}~\bibnamefont{Mikami}}, \bibinfo {author}
  {\bibfnamefont{H.}~\bibnamefont{Tokuda}},\ and\ \bibinfo {author}
  {\bibfnamefont{M.}~\bibnamefont{Watanabe}},\ }%
  \bibfield{journal}{%
  \bibinfo {journal} {Journal of Physical Chemistry B}\ }%
  \textbf{\bibinfo {volume} {113}},\ \bibinfo {pages} {10641} (\bibinfo {year}
  {2009})%
  \bibAnnoteFile{NoStop}{tsuzuki_jpcb_2009}%
\end{thebibliography}

\end{document}